\documentclass[review]{elsarticle}
\usepackage[margin=1in]{geometry}

\usepackage{titlesec}
\usepackage{adjustbox}
\titleformat{\paragraph}
{\normalfont}{\it\theparagraph}{1em}{} 
\titlespacing*{\paragraph}
{0pt}{3.25ex plus 1ex minus .2ex}{1.5ex plus .2ex}  

\usepackage{hyperref} 

\usepackage[version=3]{mhchem}
\usepackage{amsmath}
\usepackage{wasysym}
\usepackage{graphics}
\usepackage{lscape}

\journal{Earth Science Reviews}

\biboptions{authoryear}

\begin{document}

\begin{frontmatter}

\title{The Nitrogen Budget of Earth}


\author[mymainaddress]{\corref{mycorrespondingauthor}{Ben Johnson}}
\cortext[mycorrespondingauthor]{Corresponding author}
\ead{bwjohnso@uvic.ca}

\author[mymainaddress]{Colin Goldblatt}
\address[mymainaddress]{University of Victoria, Department of Earth and Ocean Sciences, 3800 Finnerty Road, Victoria, BC V8P 5C2, Canada}
\ead{czg@uvic.ca}

\begin{abstract}
 We comprehensively compile and review N content in geologic materials to calculate a new N budget for Earth. Using analyses of rocks and minerals  in conjunction with N-Ar geochemistry demonstrates that the Bulk Silicate Earth (BSE) contains $\sim7\pm4$ times present atmospheric N ($4\times10^{18}$ kg N,  or PAN), with $27\pm16\times10^{18}$ kg N. Comparison to chondritic composition, after subtracting N sequestered into the core, yields a consistent result, with BSE N between $17\pm13\times10^{18}$ kg to $31\pm24\times10^{18}$ kg N. Embedded in the chondritic comparison we calculate a N mass in Earth's core ($180\pm110~\textrm{to}~300\pm180\times10^{18}$ kg) as well as present discussion of the Moon as a proxy for the early mantle. 
 
 Significantly, our study indicates the majority of the planetary budget of N is in the solid Earth.  We suggest that the  N estimate here precludes the need for a ``missing N'' reservoir. Nitrogen-Ar systematics in mantle rocks and primary melts identify the presence of two mantle reservoirs: MORB-source like (MSL) and high-N. High-N mantle is composed of young, N-rich material subducted from the surface and identified in OIB and some xenoliths. In contrast, MSL appears to be made of old material, though a component of subducted material is evident in this reservoir as well. 
 
Taking into account N mass and isotopic character of the atmosphere and BSE, we calculate a $\delta^{15}$N value of $\sim2\permil$. This value should be used when discussing bulk Earth N isotope evolution. Additionally, our work indicates that all surface N could pass through the mantle over Earth history, and in fact the mantle may act as a long-term sink for N. Since N acts as a tracer of exchange between the atmosphere, oceans, and mantle over time, clarifying its distribution in the Earth is critical for evolutionary models concerned with Earth system evolution. We suggest that N be viewed in the same light as carbon: it has a fast, biologically mediated cycle which connects it to a slow, tectonically-controlled geologic cycle. 

\end{abstract}


\end{frontmatter}


\section{Introduction}

\label{}

\begin{figure*}[t]
\centering
\includegraphics[width=0.6\textwidth]{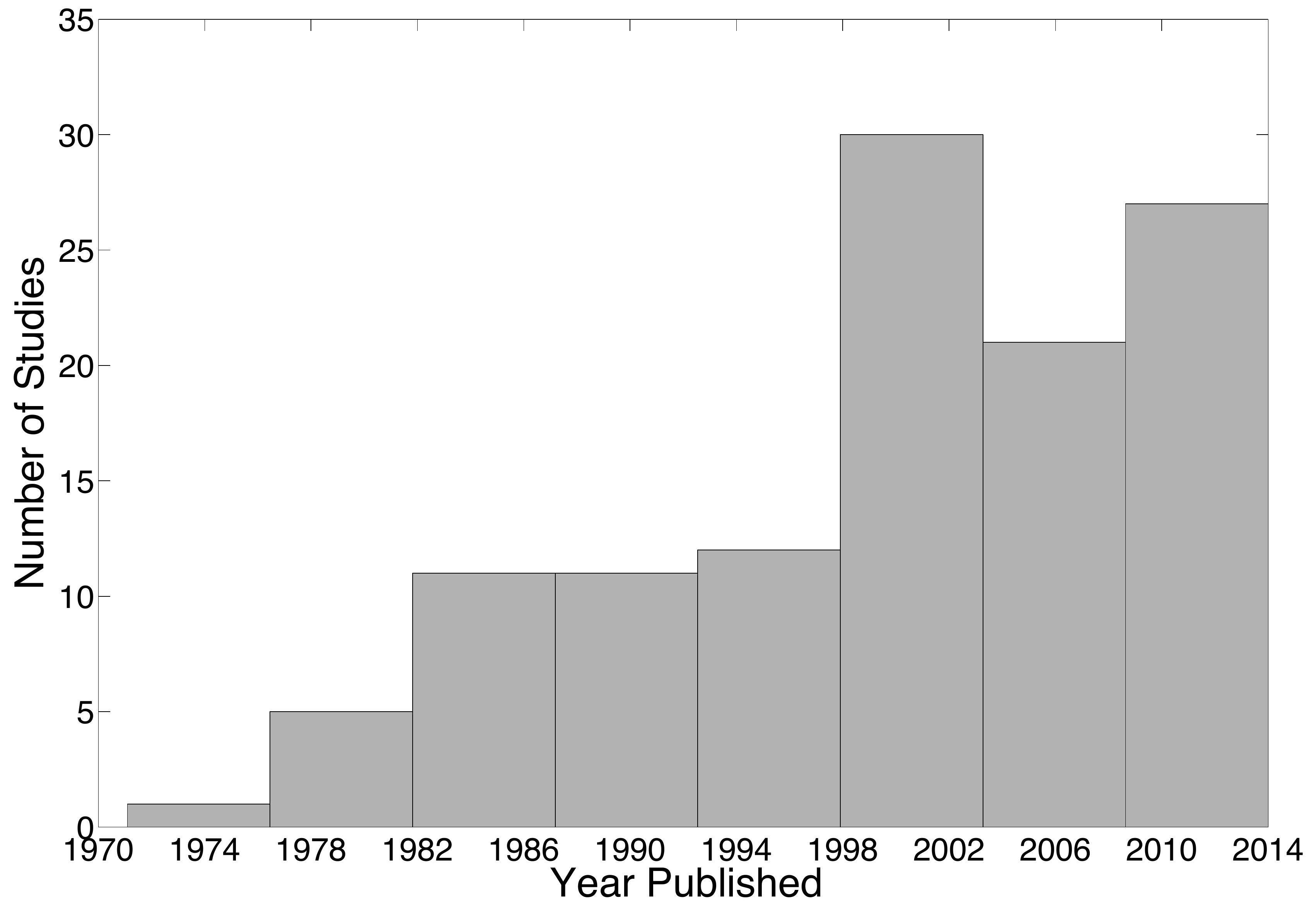}
\caption{Number of studies measuring N in geologic materials since 1975. The number of studies has increased as detection capability improves. Data produced after the mid 1990s have not been incorporated into a broad, Earth system perspective on the N cycle.} 
\label{fig:year}
\end{figure*}

Nitrogen, the fifth most common element in the solar system,  is the main component of the atmosphere, is a key nutrient for life, and has potential to be a tracer of processes linking the surface Earth to different reservoirs in the solid planet. Though N has long been known to exist geologically  in fluid inclusions or as NH$\mathrm{_4^+}$ in mineral lattices \citep[e.g.,][]{Mayne_1957}, it was thought to predominantly reside in the atmosphere and biosphere \citep{Baur_and_Wlotzka_1969}. It is now clear that N can indeed become incorporated into minerals and rocks in significant amounts and cycles over long time scales through the atmosphere, oceans, crust, and mantle. While the absolute concentration of N in rocks is low (often $\sim$1 ppm, but up to $\sim$100 or 1000 ppm), the great mass of the solid Earth compared to the atmosphere means that it has the potential to sequester large amounts of N.  A  picture of the behaviour of N in the Bulk Silicate Earth (BSE)  has begun to emerge, but necessitates a new review and synthesis of available data (Fig. \ref{fig:year}).

Similar to C \citep[e.g.,][]{Holland_1984}, N is cycled in the Earth system in two ways: a fast, biologic cycle; and a slow, geologic cycle. Descriptions of  biologic \citep[e.g.,][]{Kelly_2000} and geologic \citep[e.g.,][]{Boyd_2001,Holloway_and_Dahlgren_2002, Kerrich_et_al_2006} N cycles exist, but no adequate Earth system-wide picture of the fast and slow N cycles together is currently available. Briefly, the biologic cycle (for the modern Earth) is as follows: N$_2$ in the atmosphere dissolves in the ocean and is converted to a biologically available form by N-fixing bacteria. This process is termed N-fixation. Nitrogen-fixing bacteria are either consumed by other organisms, or release N in waste, primarily as NH$_4^+$, which is quickly oxidized to NO$_3^-$ in a bacterially-mediated process called nitrification.  The primary return flux of N to the atmosphere is via denitrification, where NO$_3^-$ is used by certain bacteria as the terminal acceptor in the electron transport chain and converted to either N$_2$ or N$_2$O. Recently, the importance of an additional reaction, anaerobic ammonium oxidation or anammox has been recognized as a return flux of N to the atmosphere. \citep[][and references therein]{Thamdrup_2012}. This is another bacterially mediated process whereby NH$_4^+$ reacts with NO$_2^-$ to produce N$_2$ and two H$_2$O molecules. 

The slow geologic cycle begins when dead organic matter sinks and settles in oceanic sediment. Organic N breaks down in the sediment via hydrolysis reactions, and converts to NH$_4^+$ \citep{Hall_1999}. Since NH$_4^+$ has the same charge and a similar ionic radius as K$^+$, it substitutes into mineral lattice sites that are normally occupied by K$^+$. Clay minerals, micas, and K-feldspars are important mineral hosts of N. Once entrained in oceanic sediments and crust, N is carried into subduction zones, where it is either volatilized and removed from the down-going plate or carried into the mantle past the subduction barrier. In general, subduction zones with high geothermal gradients favour volatilization \citep[e.g.,][]{Elkins_et_al_2006}, while cooler subduction zones favour N retention \citep[e.g.,][]{Mitchell_et_al_2010}. Volatilized N either oxidizes to N$_2$ and escapes via arc volcanism or is incorporated into intrusive igneous rocks. Nitrogen that is not returned to the surface becomes entrained in mantle circulation. Basalts at both mid-ocean ridges (MORB)  \citep{Marty_1995} and ocean islands (OIB) \citep{Mohapatra_et_al_2009} show evidence for this surface-derived N, through either positive $\delta^{15}$N values\begin{footnote} {Stable isotope notations are in per mil ($\permil$) notation, where \begin{equation} \delta^{X}\textrm{E}(\permil)=\left(\frac{^{X}\textrm{E}/^{x}\textrm{E}_{sample}}{^{X}\textrm{E}/^{x}\textrm{E}_{standard}}-1\right)*1000\end{equation} E is element of interest, $X$ is heavy isotope, $x$ is light isotope. $\delta^{13}$C standard is V-PDB and the $\delta^{15}$N standard is N$_2$ in air, which have  a $\delta^{13}$C or $\delta^{15}$N value of $0\permil$ by definition.} \end{footnote} (OIB) or correlation with radiogenic Ar (Sec. \ref{sec:atmosphere}). 
 
While the general outline of the geologic N cycle is known, in order to more fully quantify this cycle and describe changes in it over Earth history, we calculate a thorough inventory of the N on Earth.  This is a necessary step to accurately portray the Earth-system nature of the N cycle. To achieve this goal we present two approaches: a ``top-down'' and ``bottom-up'' budget estimates. The ``top-down'' approach uses the composition of planetary building blocks and analogues to bracket total Earth N content. We then subtract the amount of N in the core to estimate BSE N content. For the ``bottom-up'' approach, we compile  analyses of N in terrestrial rocks and minerals. We use these  to estimate N concentration in various reservoirs: oceanic and continental sediments, oceanic and continental crust, and the mantle. We also use observed relationships between N and Ar from basalts to estimate the mantle N content. In addition, we briefly discuss the behaviour of N in specific reservoirs. Our approach differs from past attempts by utilizing an extensive literature compilation in conjunction with new experimental results to provide a  thorough, comprehensive assessment of the N in all reservoirs of the Earth.

The structure of the paper is to first present description of the speciation and behaviour of N in the solid Earth, then a brief discussion of the data compilation used herein; this is followed by the two budget approaches, and finally a discussion of the implications of results. We present a discussion of N speciation and solubility first to serve as orientation, as N can exist as different species in the Earth depending on physical and chemical conditions. A flurry of recent experiments have elucidated many aspects of N solubility in silicate minerals \citep[e.g.,][]{Li_et_al_2013}, metal alloys \citep{Roskosz_et_al_2013}, and fluids \citep{Li_and_Keppler_2014}. 

Ultimately, we find that both approaches are mutually consistent. Chondritic comparison suggests between $17\pm13\times10^{18}$ kg to $31\pm24\times10^{18}$ kg N in the BSE ; terrestrial compilation suggests $27\pm16\times10^{18}$ kg N in the BSE. Our work indicates  not only suggests a higher N mass in the BSE than previous work \citep{Goldblatt_et_al_2009}, it arrives at approximately the same value from two independent tactics. A higher N content may have important implications for the geochemical history of N on the Earth.  In addition, our budget allows for a reassessment of the overall N-isotopic composition of the planet, which is used to track interaction between various reservoirs on the Earth. These implications are detailed in our discussion (Sec \ref{sec:discussion}).

 \begin{center}
\begin{table*}
\caption{Previous estimates for the N budget of the silicate Earth. There is significant disagreement between estimates, necessitating a more comprehensive approach. All values are $10^{18}$ kg N}
\centering
\begin{tabular}[h]{ l c c}  
\hline\\

	\bf{Reservoir}	&	\bf{Amount} &\bf{Reference}	\\[2ex] \hline
	BSE & 2.78	&\cite{Halliday_2013} \\  
	   Mantle &	5 & \cite{Marty_2012}	\\ 
	    & $\ge8.4\pm5.2$ & \cite{Goldblatt_et_al_2009} \\ 
  Continental Crust & $2.1\pm1.1$& \cite{Goldblatt_et_al_2009}\\
  &1.1 & \cite{Rudnick_and_Gao_2003, Rudnick_and_Gao_2014} \\ 
  & 1.3&\cite{Wedepohl_1995} \\
 & 14 & \cite{Delwiche_1970}\\
 Continental Sediments & 4& \cite{Delwiche_1970} \\
[1ex]
\hline

\label{tab:previous}
\end{tabular}
\end{table*}
\end{center}

\section{Nitrogen speciation in geologic materials, experimental results, and budget tools}
In this section, we first summarize which N species are found in geologic materials, highlighting silicate rocks and minerals, fluids, and Fe-metal. Secondly, we   incorporate recent experimental work to attempt to quantitatively describe N behaviour in geologic materials in response to changes in pressure, temperature, and oxygen fugacity. Thirdly, we describe the database used for subsequent budget calculation. Details pertinent to specific reservoirs will be discussed in the appropriate sections.  

\subsection{Nitrogen speciation in the solid Earth}

Nitrogen is present as a number of species in the solid Earth. The primary control on speciation is redox, with temperature, pressure, and even pH playing roles in stability and solubility.  Oxygen fugacity ($f_{O_2}$)  is  presented relative to some mineralogically controlled buffer \citep{Frost_1991}. Buffers used in this study, in order of decreasing $f_{O_2}$, are Nickel-Nickel Oxide (NiNiO), Fayalite-Magnetite-Quartz (FMQ), and Iron-W\"ustite (IW). Important N species in the solid Earth are, in order of decreasing oxidation state,  N$_2$ (fluid inclusions and degassing magmas) \citep[e.g.,][]{Marty_1995}, NH$_3$ (in reduced fluids) \citep{Li_and_Keppler_2014}, NH$_4^+$ (stably bound in mineral lattices) \citep[e.g.,][]{Itihara_and_Honma_1979}, and nitrides (e.g., FeN) \citep[e.g.,][]{Adler_and_Williams_2005}. Small differences in pH \citep{Mikhail_and_Sverjensky_2014}, especially in the mantle, may also exert some control over N speciation, though this is likely secondary when compared with $f_{O_2}$. 

There are three important reservoirs that contain the various species of N: silicate rocks and minerals, fluids and magmas, and Fe-metal. In general, N in silicate rocks and minerals is found in reduced forms, as either organic material or, more importantly for  stable geologic incorporation, as NH$_4^+$. While  there are examples of N-silicates (e.g., buddingtonite (\ce{NH4AlSi3O8}) and tobelite (\ce{(NH4, K)Al2(Si3Al)O8(OH)2})), a much more important path for N incorporation into minerals is the substitution of trace amounts of NH$_4^+$; this mechanism  is the most geologically stable way for N to be found in minerals and rocks. 
Ammonium has, depending on coordination, an ionic radius that is $<0.2~\AA$ larger than the ionic radius of K$^+$ (1.61--1.69 vs. 1.46--1.63), and  can readily substitute into K-bearing minerals \citep{Whittaker_and_Muntus_1970, Khan_and_Baur_1972} or for Na and Ca in plagioclase feldspars \citep{Honma_and_Itihara_1981}.  Indeed, K and N concentrations are correlated in sedimentary (especially metasedimentary) rocks, though this relationship is less clear in other rock types \citep[e.g.,][]{Busigny_et_al_2005b}. The source of the NH$_4^+$ can either be dead organic matter, which breaks down  into amino acids and is subsequently hydrolized during burial, or some previous inorganic source \citep{Hall_1999}. In general, N concentrations decrease with increasing metamorphic grade \citep[e.g.,][]{Haendel_et_al_1986, Bebout_and_Fogel_1992}, though the NH$_4^+$-Si bond can be quite resilient during metamorphism \citep[e.g.,][]{Pitcairn_et_al_2005,Palya_et_al_2011}. It is also possible for N to be found as  N$^{3-}$ \citep{Libourel_et_al_2003}, which can substitute for O$^{2-}$ in silicate lattices or bond with metals \citep{Roskosz_et_al_2013}.

In contrast with silicate rocks and minerals, most fluids and magmas  originating from the crust or upper mantle are oxidizing, with an $f_{O_2}$  near the  (FMQ) buffer.  At fugacity near FMQ, both natural samples \citep[e.g.,][]{Marty_1995, Nishizawa_et_al_2007} and experimental results \citep[e.g.,][]{ Libourel_et_al_2003, Li_and_Keppler_2014} show that N$_2$ is the dominant N species in magmas and fluids. At more reduced ($f_{O_2}<$FMQ) conditions, NH$_3$ becomes stable in fluids, and may even dominate in some crustal and upper mantle conditions \citep{Li_and_Keppler_2014}.

The third important reservoir for N is Fe-metal. Nitrogen is quite soluble in Fe-metal alloys at a variety of depths in the Earth \citep{Kadik_et_al_2011, Roskosz_et_al_2013}. It likely either dissolves as NH$_3$ or forms Fe-N (nitride) compounds. 
 This has important ramifications for the N distribution in the Earth. Not only could significant N be found in Earth's core, Fe-Ni metal may be present in the mantle transition zone and lower mantle \citep{Frost_and_McCammon_2008}. There might be $\le$10 wt.$\%$ N in FeNi-metal and $\le$0.5 wt.$\%$ N in  silicates in the transition zone and lower mantle \citep{Roskosz_et_al_2013}. These concentrations indicate that an enormous quantity of N are theoretically plausible in the deeper domains of the mantle. This is discussed in more detail later. 

Since N concentrations in geologic materials are usually quite low, analytical techniques present a non-trivial obstacle. A thorough discussion on this subject is provided by both \cite{Holloway_and_Dahlgren_2002} and \cite{Brauer_and_Hahne_2005}. Briefly, N can be measured by dissolution/combustion and analysis on a mass spectrometer, spectral methods, Kjeldahl extraction, or colorimetric methods. These techniques continue to evolve and improve \citep{Yokochi_and_Marty_2006, Barry_et_al_2012}, and the availability of quality N data from rocks will continue to grow. 

\subsection{Experimental results}

\begin{figure*}[]
\centering
\includegraphics[ keepaspectratio=true, width=\textwidth]{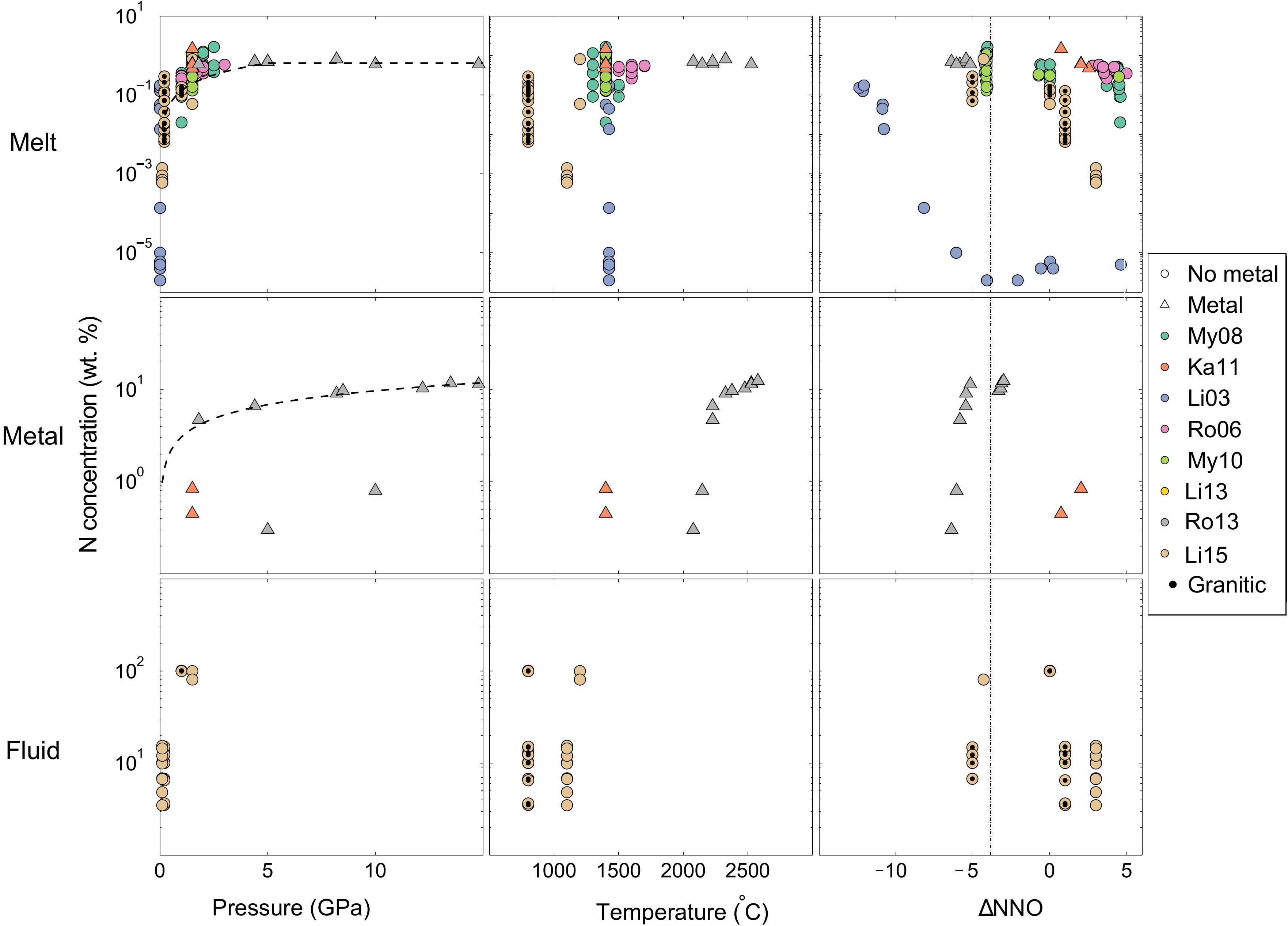}
\caption{Compilation of recent experiments measuring N solubility in silicate melts, Fe-metal, and aqueous fluids. Experiments that have silicate and Fe-metal in equilibrium ($\Delta$) and those with no metal ($\circ$) are shown.  Note log scale for N concentration. Different colours refer to specific studies: My08 \citep{Mysen_et_al_2008}, Ka11 \citep{Kadik_et_al_2011}, LI03 \citep{Libourel_et_al_2003}, Ro06 \citep{Roskosz_et_al_2006}, My10 \citep{Mysen_and_Fogel_2010}, Li13 \citep{Li_et_al_2013},  Ro13 \citep{Roskosz_et_al_2013}, and Li15 \citep{Li_et_al_2015}. All experimental runs used basaltic composition, aside from the few marked ``Granitic''.  We show concentrations as a function of pressure, temperature, and $f_{O_2}$ (relative to the NiNiO buffer) for all three phases.  Dashed lines are empirical fits to data, shown in the text (Eq. \ref{eq:Ns}-\ref{eq:Nm}). Vertical dashed line in $\Delta$NNO plots represent the IW buffer ($\Delta$NNO$-4$), below which Fe-metal is stable. While $f_{O_2}$ is the primary control on N speciation, pressure appears to be very important in solubility. } 
\label{fig:exp}
\end{figure*}

\begin{figure*}[]
\centering
\includegraphics[keepaspectratio=true, width=\textwidth]{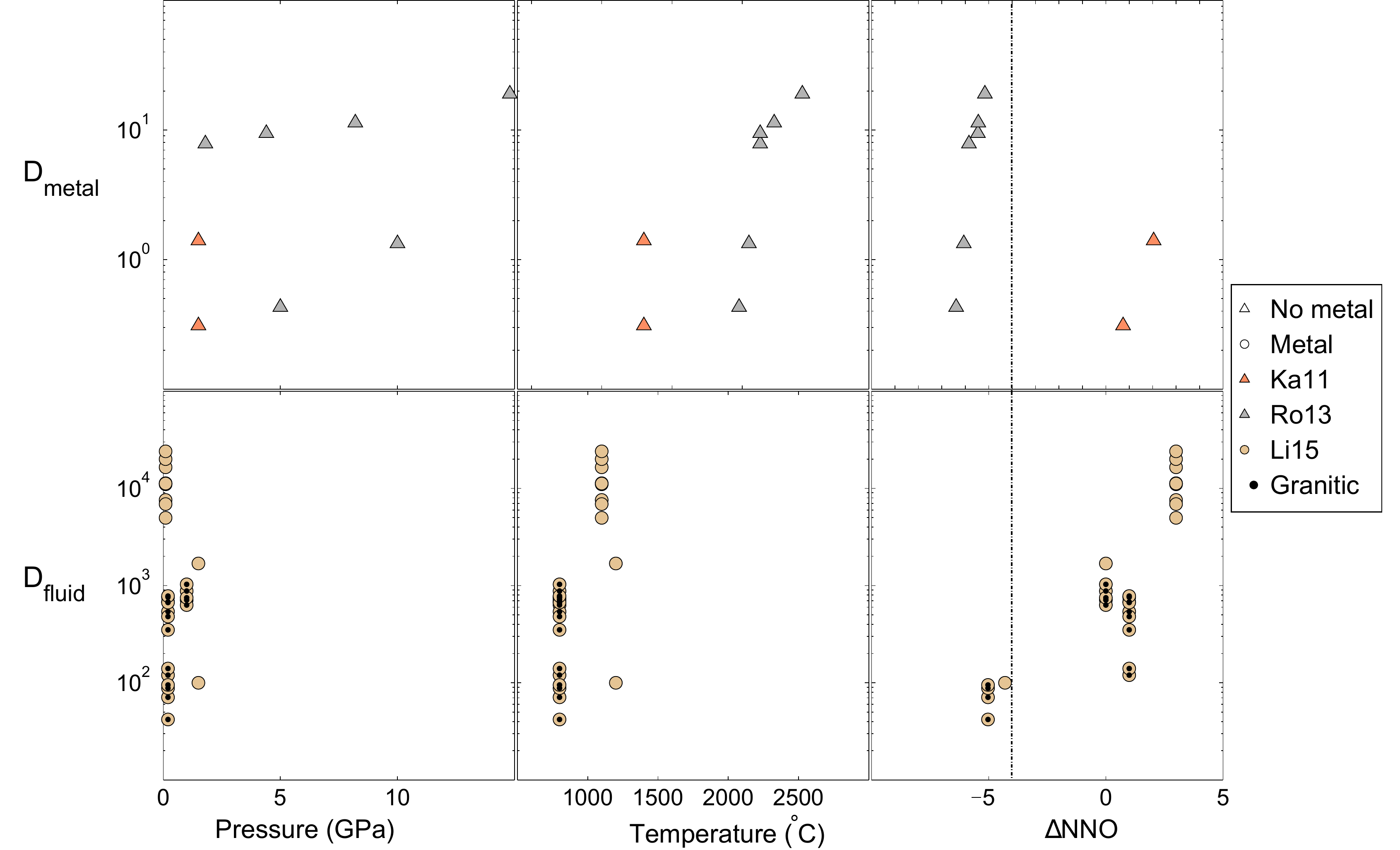}
\caption{Distribution coefficients for metal:silicate melt (top row) and fluid:silicate melt (bottom row) as a function of temperature, pressure, and $f_{O_2}$ (relative to the NiNiO buffer). Increasing pressure and temperature increases D$_\textrm{metal}$. Increasing pressure decreases  D$_\textrm{fluid}$, and temperature seems to have a negligible effect. As $f_{O_2}$ increases, N solubility in fluids increases, likely because N is present as N$_2$. References and symbols are the same as Fig. \ref{fig:exp}.} 
\label{fig:Dexp}
\end{figure*}

We have compiled experimental results to augment the discussion in the previous section and to quantitatively  describe the N solubility of geologic materials (Figs. \ref{fig:exp}-\ref{fig:Dexp}). Measurements have been made for  N in minerals \citep{Li_et_al_2013}, silicate melt \citep{Libourel_et_al_2003, Mysen_et_al_2008,  Mysen_and_Fogel_2010, Mysen_et_al_2014}, Fe-metal \citep{Kadik_et_al_2011, Roskosz_et_al_2013}, and aqueous fluids \citep{Li_and_Keppler_2014, Li_et_al_2015}.  Experimental conditions are variable (e.g., different starting materials, presence of alkalis, etc.), so at times trends are only visible when discussing single studies. Most studies use a basaltic composition for silicate components, with one using a more felsic, haplogranite material \citep{Li_et_al_2015}. In spite of these differences, however, general observations can be made from these data. Importantly, results  allow for calculation of N capacity and/or contents in poorly or unsampled reservoirs in the Earth, such as the core (Sec. \ref{sec:core}) and parts of the mantle (Sec. \ref{sec:indv}).

Pressure, temperature, and $f_{O_2}$ all have an effect on N solubility in silicate melts, Fe-metal, and aqueous fluids. A first order observation is that N concentration appears to always be higher in fluids, melts, and Fe-metal than in coexisting silicate minerals (Fig. \ref{fig:exp}). This is especially clear when the distribution coefficients (D$_\textrm{metal/fluid}=[\textrm{N}_\textrm{metal/fluid}]/[\textrm{N}_\textrm{silicate}$]) are calculated (Fig. \ref{fig:Dexp}). At all measured conditions, N prefers metal or fluid over silicates. 

Increasing pressure has noticeable effects on N solubility in silicates and metals, while the effect is less clear in fluids. Silicate N concentration increases with pressure, and, at least in the presence of Fe-metal, saturates at $0.64$ wt.$\%$ at pressures above about 5 GPa \citep{Roskosz_et_al_2013}. At lower pressures, solubility appears to follow a Henry's law relationship, given by: 
\begin{equation}  [\textrm{N}]_\textrm{S}=k_H\textrm{p} \label{eq:Ns} \end{equation}
where [N]$_\textrm{S}$ is in wt.$\%$, $k_H$ is 0.128 wt.$\%$ GPa$^{-1}$, and p is pressure (GPa). 
Concentration in Fe-metal also increases with pressure, and appears to be described by a Sievert's law equation: 
\begin{equation} [\textrm{N}]_\textrm{M}=k_s\sqrt{\textrm{p}} \label{eq:Nm} \end{equation}
where [N]$_\textrm{M}$ is in wt.$\%$, $k_s$ is an experimentally determined constant (3.06 wt.$\%$ GPa$^{-1/2}$), and p is pressure (GPa). 
The pressure effect in aqueous fluids appears to be equivalent to silicates and metal, but experiments have been done only at  lower pressures \citep{Li_et_al_2015}.

Increasing temperature results in a decrease in N content in silicate melts (Fig. \ref{fig:exp}). The effect is most clearly seen  in data  from individual studies \citep{Libourel_et_al_2003, Mysen_et_al_2008}. Higher temperatures favour formation of N$_2$, which is more easily removed from  silicate melts via extraction in fluids. Figure \ref{fig:Dexp} shows this well: higher temperature is associated with a higher D$_\textrm{fluid}$. This is partially due to the instability of N-H bonds at high temperature. Experiments done at the highest temperatures have Fe-metal in equilibrium with silicates, and since N-solubility in metal increases with increasing temperature, it is likely that N was lost from the silicates and taken up by the Fe-metal in these experiments \citep{Roskosz_et_al_2013}.

In contrast,  $f_{O_2}$ has a fairly strong effect on N solubility, and especially N partitioning between silicates and fluids (Fig. \ref{fig:Dexp}). In each experiment shown here, decreasing $f_{O_2}$ results in higher N content in silicates. This effect is less clear in metal, though these experiments were carried out at a narrower $f_{O_2}$ range, and $f_{O_2}$ must be at or below the IW buffer ($=\Delta\textrm{NNO}-4$) to even have Fe-metal stable in the experiment.  Since  oxidizing conditions promote N speciation as more fluid-mobile N$_2$, as opposed to NH$_4^+$, D$_\textrm{fluid}$ tends to decrease with decreasing $f_O{_2}$ as well. While the magnitude of the $f_{O_2}$ effect is different between different studies, the direction is the same throughout: lower $f_O{_2}$ results in higher N contents in silicates. 

There are also some measurements of N-contents in minerals directly. We utilize equations, described by \cite{Li_et_al_2013}, of N solubility experimental results for olivine, pyroxene, and melt (in the absence of Fe-metal) to guide both estimates of N concentration and distribution coefficients (described below) between minerals and melt in poorly sampled reservoirs: 
\begin{equation} \textrm{Olivine}:~\textrm{log}_{10}\textrm{~[N]} = 2.15-\frac{6.8\times10^3}{\textrm{T}}+0.27\textrm{P}-0.43\Delta\textrm{NiNiO} \label{eq:ol}; ~\textrm{r}^2=0.79 \end{equation}
 \begin{equation}\textrm{Pyroxene}:~\textrm{log}_{10}\textrm{~[N]} = 6.48-\frac{8.7\times10^3}{\textrm{T}}+0.086\textrm{P}-0.122\Delta\textrm{NiNiO}; \label{eq:pyx} ~\textrm{r}^2=0.64 \end{equation} 
 \begin{equation}\textrm{Melt}:~\textrm{log}_{10}\textrm{~[N]} = 0.92-\frac{3.50\times10^3}{\textrm{T}}+0.4\textrm{P}-0.083\Delta\textrm{IW}; \label{eq:melt}~\textrm{r}^2=0.70  \end{equation} 
The above equations have  temperature (T) in K, pressure (P) in GPa, $\Delta\textrm{NiNiO}$ or $\Delta\textrm{IW}$  is the $f_{O_2}$ relative to the NiNiO or IW buffer, and [N] is in ppm.  At appropriate conditions, concentrations of up to 100 ppm may be possible in the lowermost upper mantle \citep{Li_et_al_2013}, which means the upper mantle may have the capacity to sequester $\sim80\times10^{18}-200\times10^{18}$ kg  N, which is $20-50$ times PAN.

The last tool based on experiments we utilize is  measured or inferred partition coefficients \newline (K$_\textrm{D}=$[Element]$_\textrm{mineral}$/[Element]$_\textrm{melt}$); these are often used in conjunction with  an equation linking partition coefficients to degree of partial melting  \citep{Rollinson_1993}. 
\begin{equation} \frac{[\textrm{C}_\textrm{L}]}{ [\textrm{C}_\textrm{o}]} = \frac{1}{\textrm{K}_\textrm{D}+\textrm{F}(1-\textrm{K}_\textrm{D})} \label{eq:partial} \end{equation}
[C$_\textrm{o}$] is element concentration in source and [C$_\textrm{L}$] is concentration in melt,  and F is degree of partial melting.
Note that this equation is for batch (equilibrium) melting, which means that melt formed equilibrates with residual solids. We assume that any melt much reach a critical threshold ($\sim1-10\%$) before extraction from the source rock, and  prior to extraction it would have time to equilibrate fully with residual solids.

\subsection{Database of geologic N measurements} 
We have compiled all of the available, published measurements of N concentration and $\delta^{15}$N values of geologic materials. Where they exist, we also include in the database $\delta^{13}$C, age of sample, Ar-isotope ratios and abundance, and concentrations of elements that behave similarly to NH$_4^+$, including K$_2$O, Rb, Lu, and Yb. The complete database is available in the supplementary material, which is organized by both rock names, as given in the original publications, and our interpreted geologic settings, 

While rock names follow standard naming procedure, we also categorize data based on geologic setting. Unmetamorphosed samples are labeled as oceanic sediments (OS), oceanic lithosphere (OL), continental sediments (CS), and continental lithosphere (CL). Altered reservoirs (i.e., metamorphosed at $\textrm{T}<300~^{\circ}$C) are prefixed with `A'; those metamorphosed at $\textrm{T}>300~^{\circ}\textrm{C}$ are prefixed with `M'. Data for the mantle are from diamonds (D) and xenoliths (X). We also discuss mid-ocean ridge basalts (MORB) and ocean island basalts (OIB). These reservoirs will be addressed individually in following sections. 

Nitrogen concentration from most reservoirs are log-normally distributed. To calculate N mass in a given reservoir, we will generally use the product of the log-normal mean of N concentration and mass of that reservoir. As sample size is often low, we calculate maximum likelihood estimator parameters $\hat{\mu} ~\textrm{and}~ \hat{\sigma}^2$, which are the mean and variance of the natural log of concentration, respectively \citep{Limpert_et_al_2001}.     
\begin{equation}\hat{\mu}=\frac{\sum_i\textrm{ln[N]}_i}{n}\end{equation}
\begin{equation}\hat{\sigma}^2=\frac{\sum_i(\textrm{ln[N]}_i-\hat{\mu})^2}{n}\end{equation}
Where n is the number of samples. 
These parameters are then used to estimate the mean ($\mu$) and standard deviation ($\sigma$) of the total population: 
\begin{equation}\mu\simeq \bar{x}=e^{\hat{\mu}+\frac{\hat{\sigma}^2}{2}} \end{equation}
\begin{equation} \sigma=\sqrt{(e^{\hat{\sigma}^2}-1)e^{2\hat{\mu}+\hat{\sigma}^2}} \end{equation}
Unless otherwise specified, all errors given are standard error of the mean: 
\begin{equation}\textrm{SE}\bar{x}=\frac{\sigma}{\sqrt{n}} \end{equation}

\section{``Top-down'' Budget: Accretion through Core formation}\label{sec:topdown}
In this section, we estimate the N budget of the BSE by comparing the Earth to other inner solar system bodies. The atmosphere of Venus hints that there is more N in the Earth than is found in its atmosphere alone. We  bracket mass of N delivered to Earth during accretion by comparison to chondritic compositions. From this, we subtract N sequestered into the core to estimate the remainder in the BSE and atmosphere. While this  model is dependent on the N content of accretionary material, we find that it is in reasonable agreement with our terrestrial-based budget, presented in Section \ref{sec:bottomup}. In addition, the N content of the Moon is calculated, as this may provide some constraints on the composition of the early, but post-core formation, mantle. 

\subsection{Initial N composition and planetary comparison: missing N?} \label{sec:initialN}

Some  motivation for this study comes from comparison of the Earth to extraterrestrial bodies: meteorites and Venus. Undifferentiated meteorites are leftover remnants from the early history of the Solar System, and are often used as proxies for the bulk composition of the protoplanetary disk. Venus  is thought to have had a  similar initial volatile composition as the Earth \citep{Ringwood_and_Anderson_1977, Lecuyer_et_al_2000, Chassefiere_et_al_2012}. Comparison to both meteorites and Venus suggest that the Earth should have much more N than is found in the present atmosphere; by extension, we posit that the atmosphere is not the major N reservoir on Earth.

 We address Venus first. The Venusian atmosphere contains $3.5\pm0.8\%$ N$_2$, with the remainder composed of predominately ($96.5\%$) CO$_2$  \citep{van_Zahn_et_al_1983}. We  calculate the mass of N (M$_{\textrm{N}_2}$) in the atmosphere by using the following equation: 

\begin{equation}\textrm{M}_{\textrm{N}_2}=\frac{\textrm{m}_{\textrm{N}_2}}{\textrm{m}_{\textrm{a}}}\cdot x_{\textrm{N}_2}\cdot \frac{4\pi\textrm{r}^2\textrm{p}}{\textrm{g}}
\label{eq:atmosphere}\end{equation}

where m$_{\textrm{N}_2}$ and m$_{\textrm{a}}$ are molar masses of N$_2$ (0.028 kg mol$^{-1}$) and Venus' atmosphere (0.04344 kg mol$^{-1}$); $x_{\textrm{N}_2}$ is the mixing ratio of N$_2$  (0.035); r is the radius of Venus ($6.052\times10^6$ m); p is surface pressure ($9.2\times10^6$ Pa); and g is acceleration due to gravity (8.87 m s$^{-2}$). The resulting N content of Venus' atmosphere is 11$\times10^{18}$ kg N. When normalized to planetary mass, Venus' atmosphere has 3.4 times the mass of N in Earth's atmosphere. Given similar initial volatile composition, Earth should have substantial N in non-atmospheric reservoirs. Curiously, the amount of C in the Venusian atmosphere (as CO$_2$) is nearly identical to the amount of C in carbonate rocks on Earth \citep{Taylor_1992, Berner_1998, Lecuyer_et_al_2000}. If a similar mass balance exists for N, then a substantial amount of N must be in geologic reservoirs on Earth.

\begin{figure*}[t]
\centering
\includegraphics[keepaspectratio=true, height=3.25in]{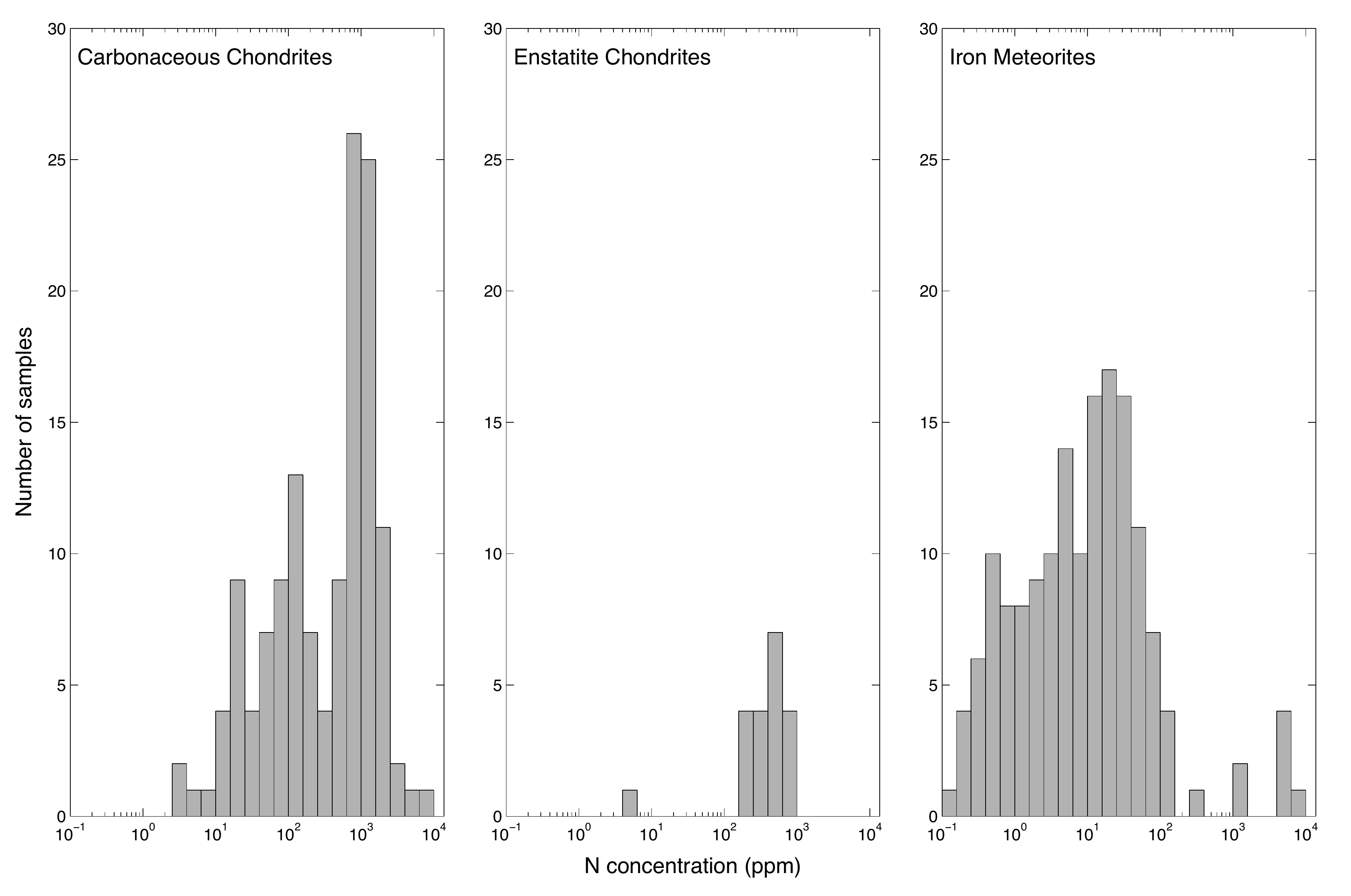}
\caption{Nitrogen concentration in carbonaceous chondrites (CC), enstatite chondrites (EC), and iron meteorites. Nitrogen content of both CC ($1235\pm440$ ppm) and EC ($605\pm2-6$ ppm) are significant, and suggest many atmospheric masses of N were delivered to the Earth during accretion. Iron meteorites are presented as a proxy for N content of the core ($140\pm10$ ppm, Sec. \ref{sec:core}  See supplemental information for data table.} 
\label{fig:meteorites}
\end{figure*}

Whilst the exact nature and composition of planetary accretionary bodies are a matter of debate, \citep[e.g.,][]{Marty_2012, Halliday_2013}, some combination of chondrite-like material accreted to form the rocky planets, including Earth. The volatile content of these bodies is thought to have decreased with distance from the Sun, though the feeding zones of growing planets may be substantial \citep{Kaib_and_Cowan_2015}. We bracket terrestrial N content by using volatile-poor enstatite chondrites (EC) and volatile-rich carbonaceous chondrites  (CC) as analogs for possible volatile delivery material. Note we are not attempting to find a ``perfect fit'' meteorite to explain terrestrial volatiles, but simply providing some context for how much N should be present in the planet.  

To utilize N contents of  chondrite proxies, we follow the approach of \cite{Marty_2012} for both CC and EC. Marty compared two chondrites' (Orguiel and Murchison) volatile abundances to a calculated volatile budget for the Bulk Earth (BE).  These specific meteorites were chosen as they are  primitive in composition and have experienced low grades of metamorphism. We include both a broader suite of CC and EC  analyses. Both chondrite types have substantial N content: EC have an average N concentration of $605\pm206$ ppm and CC $1235\pm440$ ppm (Fig. \ref{fig:meteorites}).

Non-N volatile elements  (e.g., C, H$_2$O, halogens) appear to be depleted  in the Earth relative to chondritic concentration \citep{Marty_2012}.  These volatiles are expected to have negligible concentrations in the core, which is likely not the case for N, as discussed in the next section. Therefore, we assume that the abundance calculated by Marty for the BSE plus atmosphere accounts for the total abundance, and differences from chondritic values are due to processes during accretion/delivery. Note that we exclude  Xe from this comparison, as it  is more depleted than other volatiles, and requires explanation beyond the scope of this paper \cite[e.g.,][]{Pujol_et_al_2011}. 

Overall, we show that the Earth appears to be depleted by about an order of magnitude compared to chondritic values (Table \ref{tab:volab}), which is consistent with \cite{Marty_2012}.  Using only Orguiel (CI-chondrite) and Murchison (CM-chondrite)  suggests terrestrial volatiles are $2.48\pm0.3\%$ as abundant as they are in CI/CM-chondrites. Incorporating analyses of a broader suite of CC gives an indistinguishable volatile abundance pattern, with terrestrial volatiles  $2.75\pm0.2\%$ as abundant as CC. The latter  is adopted here. Enstatite comparison yields less consistent results, though are within an order of magnitude ($9.2\pm0.1\%$). This value was calculated without Ne-abundance, as this appears to be distinct from other volatiles (Table \ref{tab:volab}).

If N behaved similarly to other volatiles during accretion/delivery, the abundance values can be used in concert with N concentrations of CC and EC to calculate BE N. Multiplying CC N content ($1235\pm440$ ppm) by BE/CC abundance ($2.75\pm0.2\%$) gives BE N  mass of $204\pm75\times10^{18}$ kg N; the same calculation with EC N content ($605\pm206$ ppm) and EC/BE abundance of $9.2\pm0.1\%$ gives BE N mass of $330\pm120\times10^{18}$ kg N. These masses are equivalent to a BE N concentration of $34\pm12$ ppm and $55\pm20$ ppm, respectively. For comparison, both N mass estimates  are two orders of magnitude greater than the mass of N in the present atmosphere ($4\times10^{18}$ kg). 

While the preceding approach is appropriate if N had similar behaviour to noble gases, water, and C during accretion, it is possible that N may have existed in reduced forms in the protoplanetary disk. Ammonia in comets is well known \citep{Oro_1961}, and recent identification of NH$_3$ as inclusions in primitive chondrites indicates that reduced N was also present in the chondrite-forming region of the solar system \citep{Harries_et_al_2015}. If N was found as NH$_4^+$ in significant quantities in the Earth-forming region, it may have behaved more like K or Rb than noble gases. We note that NH$_3$ was likely found mostly in ices, and its behaviour would be quite different than NH$_4^+$ substituting into silicate lattices or Fe-metal. The following discussion assumes N was found as NH$_4^+$ in the Earth-forming region. 

Estimates of BE N based on K and Rb content of CC and EC are higher than noble gas constraints (Table \ref{tab:krbab}). EC have higher K (770 ppm) and Rb (2.5 ppm) concentrations  than CC ([K]=400 ppm, [Rb]=1.7 ppm) \citep{Wasson_and_Kallemeyn_1988}. The Bulk Earth (BE) has 280 ppm K \citep{Arevalo_et_al_2009} and 0.6 ppm Rb \citep{Palme_and_Oneill_2014}. These abundances suggest the Earth has about 1/3 as much K or Rb as chondrites. If N behaved like K or Rb, it would have a very large mass in the BE of between $870-5200\times10^{18}$ kg N (Table \ref{tab:initialcoreN}). Since N is likely more volatile than K and Rb, this  provides a strict upper limit on N abundance in the Earth. For the remainder of the paper we adopt the CC- and EC-volatile based proxy, but do not exclude N behaving somewhere in between more volatile elements and K or Rb during planetary formation. 

It should be noted that neither class of chondrite appear to fully satisfy the isotope composition of volatile elements on Earth. Both EC \citep{Grady_et_al_1986} and CC \citep{Pearson_et_al_2006} have  $\delta^{13}$C values similar to the mantle value of $-5\permil$. A significant problem with EC as proxy for volatile delivery is that they have negligible water content, and therefore very low H. In contrast, CC  are more water-rich and have $\delta$D values are more or less consistent with at least the surface reservoirs of Earth \citep{Marty_2012}. The $\delta^{15}$N values of the mantle ($-35\permil~\textrm{to}~-5\permil$) match more closely with EC, $\sim-35\permil$ \citep{Grady_et_al_1986}, than with CC, which are variable, but consistently positive \citep{Pearson_et_al_2006}.

\begin{table*}[t]

\caption{Estimated volatile concentrations for C, H$_2$O, Ne, Ar, and Kr in chondrites after \cite{Marty_2012}, used to estimate volatile retention during accretion. Shown are concentrations in  CI-CM chondrite (CI/CM), the most primitive carbonaceous chondrites, analyses from all classes of carbonaceous chondrites (CC), enstatite chondrite (EC), and bulk Earth (BE, which is BSE plus atmosphere). We do not include Xe, which is depleted compared to chondrites and other volatiles and requires additional explanation beyond the scope of this paper. Concentrations are in mol g$^{-1}$, and abundances are shown in percent. References are indicated with superscripts. Errors for concentrations are $1\sigma$. Abundance errors are shown as SE$_{\bar{x}}$, with $1\sigma$ values given in parentheses. We calculated SE$_{\bar{x}}$ based on $1\sigma$ values and number of analyses for each volatile. SE$_{\bar{x}}$ are used in subsequent calculations. } 
\centering
\begin{adjustbox}{max width=\textwidth}

\begin{tabular}{l c c c c  c c c}   \hline \\
\bf{Species} & \bf{CI/CM$^1$ (mol g$^{-1}$)} & \bf{CC$^{2-3}$ (mol g$^{-1}$)}& \bf{EC$^{4-6}$ (mol g$^{-1}$)} & \bf{BE$^1$ (mol g$^{-1}$)}&  \bf{BSE/CI-CM ($\%$)} & \bf{BSE/CC ($\%$)}&\bf{BSE/EC ($\%$)}\\ \hline \\ 

$^{12}$C &  $2.00\pm0.2\times10^{-3}$ & $2.23\pm2.2\times10^{-3}$&$3.75\pm0.4\times10^{-4}$ & $4.38\pm1.7\times10^{-5}$ & $2.19\pm0.5~(1.4) $&$1.96\pm0.2~(2.0)$  &$11.7\pm2.3~(7.4)$\\ [1ex]

H$_2$O & $5.50\pm0.9\times10^{-3}$  & $7.50\pm1.8\times10^{-3}$ &--& $1.50\pm0.7\times10^{-4}$&$2.74\pm0.7~(1.9)$& $2.00\pm0.3~(1.4)$   &--\\ [1ex]

$^{22}$Ne & $2.68\pm0.2\times10^{-12}$ &$3.49\pm0.5\times10^{-12}$&$4.28\pm0.2\times10^{-12}$  &$2.66\pm0.02\times10^{-14}$ & $0.99\pm0.2~(0.5)$&$0.76\pm0.1~(0.8)$  &$0.620\pm0.1~(0.8)$ \\ [1ex]

$^{36}$Ar& $4.51\pm0.1\times10^{-11}$ &$3.36\pm0.8\times10^{-11}$& $2.22\pm0.9\times10^{-11}$ & $1.01\pm0.03\times10^{-12}$&$2.24\pm0.7~(1.8)$& $3.01\pm0.5~(4.4)$&$4.55\pm1.0~(8.0)$\\[1ex]

$^{84}$Kr& $4.98\pm0.1\times10^{-13}$& $1.23\pm0.3\times10^{-13}$&$1.06\pm0.2\times10^{-13}$  & $2.10\pm0.07\times10^{-14}$&$4.22\pm1.3~(3.4)$& $6.03\pm1.0~(8.3)$& $19.8\pm3.1~(24)$\\[1ex]
\emph{Average abundance}&&&&&$2.48\pm0.3~(1.0)$&$2.75\pm0.2~(2.0)$& $9.2\pm0.1~(5.3)$\\[2ex]

\hline
\multicolumn{6}{c}{$^1$ \cite{Marty_2012}, $^2$  \cite{Mazor_et_al_1970} $^3$, \cite{Bogard_et_al_1971}, $^4$ \cite{Crabb_and_Anders_1981}, $^5$\cite{Patzer_and_Schultz_2002}, $^6$\cite{Grady_and_Wright_2003}} 
\end{tabular} \label{tab:volab}
\end{adjustbox}
\end{table*}

\begin{table*}[t]

\caption{Concentrations of K and Rb in carbonaceous chondrites (CC) and enstatite chondrites (EC), compared to their abundance in the BSE (BE). If N were present in the solar nebula as NH$^+_4$, it may behave more similarly to these alkali elements than to noble gases. It is likely, however, that N would be more volatile than either K or Rb, so these represent upper limit estimates for N abundance in the BSE, compared to chondrites. } 
\centering
\begin{adjustbox}{max width=\textwidth}
\begin{tabular}{l c c c c c} \hline
\bf{Species} & \bf{CC (ppm)} & \bf{EC (ppm)} & \bf{BE (ppm)} & \bf{BE/CC ($\%$)} & \bf{BE/EC ($\%$)} \\[2ex] 
Rb & 1.7 & 2.5 & 0.6 & 35 &24\\
K& 400 & 770 & 280 & 70 & 36\\
\hline
\end{tabular} \label{tab:krbab}
\end{adjustbox}
\end{table*}

\begin{table*}[t]

\caption{Total Earth, core, and BSE N masses based on abundances (noble gas and K or Rb) calculated above and other proxies. Calculations from volatile and K or Rb proxies use distribution coefficients between silicate and Fe-metal at pressures and temperatures appropriate for core formation \citep{Roskosz_et_al_2013}. Details are presented in the text. Additional core N estimates are obtained from  thermodynamic calculation \citep{Zhang_and_Yin_2012} and analysis of iron meteorites \citep{Grady_and_Wright_2003}. ``Atm'' is the current atmosphere ($4\times10^{18}$ kg N). All values are in $10^{18}$ kg N. We use the CC- and EC-volatile estimates in the remainder of the text, and these are shown in bold. Errors are SE$_{\bar{x}}$} 
\centering
\begin{adjustbox}{max width=\textwidth}
\begin{tabular}{l c c c c c c c} \hline
\bf{Proxy} & \bf{Bulk Earth N} & \bf{Bulk Earth N (ppm)}& \bf{Core N mass } & \bf{Core N (ppm)}& \bf{BSE+Atm} & \bf{BSE only } & \bf{BSE only (ppm)} \\[1ex] 
\bf{CC-volatile} &$\bf{204\pm75}$ &$\bf{34\pm12}$& $\bf{180\pm110}$ &$\bf{102\pm63}$& $\bf{21\pm17}$ & $\bf{17\pm13}$ &$\bf{4.1\pm3.1}$\\
\bf{EC-volatile}& $\bf{330\pm120}$ &$\bf{55\pm20}$& $\bf{300\pm180}$ &$\bf{165\pm100}$& $\bf{35\pm28}$ & $\bf{31\pm24}$&$\bf{7.3\pm5.6}$\\
K-CC&$5200\pm1850$  &$864\pm310$& $4600\pm3500$ &$2580\pm2000$&$530\pm400$  &$526\pm396$ & $128\pm116$\\
Rb-CC&$2600\pm880$  &$430\pm150$&  $2300\pm1800$&$1300\pm1000$&  $270\pm250$& $255\pm246$& $64\pm58$\\
K-EC& $1300\pm500$ &$220\pm74$&$1100\pm900$  &$650\pm500$& $140\pm125$ &  $136\pm121$&$32\pm29$\\
Rb-EC& $870\pm300$ &$145\pm50$ & $780\pm600$&$430\pm330$ &$90\pm80$ & $86\pm76$  &$19\pm0.8$\\
Iron Meteorite&& &$250\pm20$&$140\pm10$ & & &\\
Thermodynamic caluclation&&  &$1.8\pm0.2$& &&&\\ [1ex]

\hline
\end{tabular} \label{tab:initialcoreN}
\end{adjustbox}
\end{table*}

\subsection{Core Formation, N sequestration, and remaining BSE N content} \label{sec:core}

Now that we have established some estimates for initial N content, the next step is to model N behaviour during core formation; some N was likely incorporated into the core. 
 Core formation occurred as gravitational separation of Fe, Ni, and additional elements from silicates during accretion. Nitrogen is siderophile (soluble in metal-Fe) under reducing conditions, allowing large quantities of N to be scavenged during core formation.  Because the core is geochemically isolated from the BSE \citep{Halliday_2004}, any scavenged N is effectively removed from the rest of the planet. It is therefore important to constrain how much N is in the core, which will be subtracted from a chondritic starting composition. 

There are several types of constraints  provided (Table \ref{tab:initialcoreN}). The first is  N measurements from iron meteorites, which are derived from cores of planetesimals formed early in the solar system's history \citep{Grady_and_Wright_2003}. While variable, these meteorites have an average N content of 138$\pm12$ ppm (Fig. \ref{fig:meteorites}), mostly contained in the mineral taenite (Fe$_{0.8}$Ni$_{0.2}$).  If iron meteorites are a good proxy for the core, it contains $250\pm20\times10^{18}$~kg.  
Secondly, there are calculations, based on thermodynamic properties, indicating the partition coefficient  between liquid iron and silicate melt (K$_\textrm{D}$=[N]$_{metal}$/[N]$_{silicate}$) of 1.8$\pm0.2$. This suggests 0.001 wt$\%$ N  in the core, for a N content  of $1.8\pm0.2\times10^{18}$ kg  \citep{Zhang_and_Yin_2012}. This estimate matches experimental work done at low pressures \citep[e.g.,][]{Kadik_et_al_2011}, but does not agree with experimental work done at higher pressures appropriate for core formation conditions. 

The third, and preferred, type of constraint uses our calculated CC- or EC-volatile proxies for BE N content in concert with experimental measurements of K$_\textrm{D}$ under high pressure ($5-20$ GPa). Measured  K$_\textrm{D}$  is $20\pm10$ \citep{Roskosz_et_al_2013}. The N concentration of the core can be calculated by using the following two relationships:
\begin{equation}\textrm{N}_\textrm{t}=\textrm{N}_\textrm{c}+\textrm{N}_\textrm{b} \end{equation}
\begin{equation}\textrm{N}_\textrm{t}=[\textrm{N}_\textrm{c}]\textrm{M}_\textrm{c}+[\textrm{N}_\textrm{b}]\textrm{M}_\textrm{b} \end{equation}
where M is mass, N without brackets is N mass, N in brackets is concentration, and subscripts are t for total Earth, c for core, and b for BSE. Mass of the core is $1.8\times10^{24}$ kg and mass of the BSE is $4.2\times10^{24}$ kg. Taking $\textrm{K}_\textrm{D}=[\textrm{N}_\textrm{c}]/[\textrm{N}_\textrm{b}]$, we find 
\begin{equation}[\textrm{N}_\textrm{c}]=\frac{\textrm{N}_\textrm{t}} {\textrm{M}_\textrm{c}+\frac{\textrm{M}_\textrm{b}}{\textrm{K}_\textrm{D}}}\end{equation}
A partition coefficient of $20\pm10$ and  bulk Earth N mass that is either CC-like ($204\pm75\times10^{18}$ kg) or EC-like ($330\pm120\times10^{18}$ kg)  suggests  $180\pm110\times10^{18}$ or $300\pm180\times10^{18}$ kg N is in the core. These values are very similar to iron meteorites, suggesting they are a good proxy for core composition. Were the volatile concentration based on K-Rb, not noble gases, the  N inventory would be $780-4600\times10^{18}$ kg. Importantly, all proxies indicate  that if N were present in the Earth during core formation, the majority of it is sequestered into the core. This may have had an isotopic effect on the N remaining in the BSE, though it may have been minimal due to the high temperature. No measurements of this fractionation have been made, to our knowledge.

We can estimate N remaining in the BSE and atmosphere by subtracting core N mass from the total Earth. This leaves N masses of $21\pm17\times10^{18}$ kg  and $35\pm28\times10^{18}$ kg  remaining in the BSE and atmosphere for CC-like and EC-like models, respectively. Further subtracting the amount in the modern atmosphere ($4\times10^{18}$ kg N), suggests  between $17\pm13\times10^{18}~\textrm{kg and}~31\pm24\times10^{18}$ kg N ($4.1\pm3.1$ to $7.3\pm5.6$ ppm) reside in the BSE. These estimates are higher than previous work for BSE N content, and serve as a useful comparison for the terrestrial-based, literature compilation budget presented in Section \ref{sec:bottomup}.

\subsection{A Lunar analogue for the Early Mantle?} 
The Moon formed after a Mars-size proto-planet (Theia) collided obliquely with a Venus-size proto-Earth (Tellus) at the end of planetary accretion \citep{Hartmann_and_Davis_1975},  marking the end of the so-called Chaotian Eon and the start of ``Earth'' history \emph{sensu stricto} \citep{Goldblatt_et_al_2010}. The density and composition of the Moon indicates that it formed after core-mantle differentiation on Earth. In addition, the identical O-isotope composition \citep{Wiechert_et_al_2001} of the Earth-Moon system indicates that  the Moon-forming impact was energetic enough to homogenize the Earth and its impactor, Theia. Hence, Lunar rocks may sample the very early Earth mantle.

The N content of Lunar rocks can be used to estimate  Lunar mantle, and therefore early Earth mantle, N concentrations. There are a few measurements of N in Lunar rocks, including basaltic glasses  (3 ppm), basalts (0.7 ppm), and anorthosites (1.5 ppm) \citep{Mathew_and_Marti_2001}. We  use the concentration from basalt glasses as is done for terrestrial basalts (Section \ref {sec:TheMantle}), as these are most quickly quenched and have experienced the least amount of degassing. Lunar glasses appear to be petrogenetically related to Lunar mare basalts \citep{Mathew_and_Marti_2001}, which have relatively well constrained melting conditions.  These basalts  are the result of partial melting of 1-10$\%$ at a depth of 400 km (2.85 GPa, lower pressure than the equivalent depth on Earth due to the smaller Lunar mass) \citep{Shearer_and_Papike_1999} and temperatures of $1125~^{\circ}$C \citep{Marty_et_al_2003}.  Oxygen fugacity is between IW-2 and IW-4  \citep{Marty_et_al_2003}. By comparison, terrestrial mid-ocean ridge basalts (MORB) formed at much more oxidizing conditions of IW+6 \citep{Frost_and_McCammon_2008}. 

\begin{table*}

\caption{Estimates of Lunar N content in ppm, shown as a function of $f_{O_2}$. Hypothetical mineral and melt N concentrations for conditions of basalt melting are calculated using Eq.  \ref{eq:ol}, \ref{eq:pyx} \ref{eq:melt}. These are used to calculate K$_\textrm{D}$ for a source rock that is $60\%$ olivine and $40\%$ pyroxene. Source concentration based on measurements of  glass (3 ppm) are calculated with  Eq. \ref{eq:partial}.}
\centering
\begin{tabular}[h]{l c c }
\hline \\[-2ex]
& \multicolumn{2}{l}{ \bf{ N concentration (ppm) }}\\
     \bf{Mineral (Modal $\%$)}         &$f_{O_2}=$IW-2  & $f_{O_2}=$IW-4\\
Olivine ($60\%$) & 11 &  82 \\
Pyroxene ($40\%$) &  45 &  77 \\
Melt &  31,000 & 45,000\\   [1ex]
 Bulk K$_\textrm{D}$ &$8\times10^{-4}$&$1.8\times10^{-3}$\\[1ex]
  & \multicolumn{2}{c}{\bf{Expected source N (ppm)}}\\[-1ex]
    & \multicolumn{2}{c}{\bf{at bulk K$_\textrm{D}$ = ($f_{O_2}$ as above)}}\\[0.2ex]
    \bf{Percent Partial Melt}   & $\bf{8\times10^{-4}}$&$\bf{1.8\times10^{-3}}$\\
    $1\%$ &0.03 & 0.04\\
    $10\%$ &0.3&0.3\\ [1ex]
    \bf{Average source rock concentration}& \multicolumn{2}{c}{$0.18\pm0.15$}\\
\hline
\label{tab:moon}
\end{tabular}
\end{table*}

To use these data to calculate N concentration in the Lunar mantle, we calculate a hypothetical K$_\textrm{D}$ based on a  basalt-source rock (peridotite) at the given T ($1125~^{\circ}$C), P (2.85 GPa), and $f_{O_2}$ (IW-2 to IW-4) conditions (Table \ref{tab:moon}).  First, N solubility in olivine, pyroxene, and melt are calculated using Eqs. \ref{eq:ol}, \ref{eq:pyx}, and \ref{eq:melt}. Next, bulk K$_\textrm{D}$ is calculated for a source rock is assumed to be $60\%$ olivine and $40\%$ pyroxene. These K$_\textrm{D}$ values, along with percent partial melt, are used in Eq. \ref{eq:partial} to calculate N concentration in basalt glass source region, given glass concentration of 3 ppm (Table \ref{tab:moon}). Thus we determine the average source N concentration for the Lunar mantle is  $0.18\pm0.15$ ppm. 

This calculated concentration is lower than what is predicted from the chondritic model, but is similar to analyses of terrestrial xenoliths ($0.28\pm0.25$ ppm, Sec. \ref{sec:TheMantle}). The lunar value is interpreted as a lower limit for the N concentration of the early Earth mantle, as there may well have been substantial N loss during moon formation. Some N was likely volatilized and lost to space during the moon-forming impact, and later by degassing from a lunar magma-ocean. Although degassing from a magma ocean is possible, we note that under the reducing conditions thought to characterize the lunar mantle, a significant proportion of N present in melts is NH$_4^+$ \citep{Libourel_et_al_2003, Mysen_et_al_2008}. Ammonium  can bond with Si-chains in a melt, and may be retained to a higher degree than N$_2$, which dissolves in magmas by filling spaces in between Si-chains. This behaviour of ``chemically solubility'' or ``physical solubility'' for NH$_4^+$ and N$_2$ might promote retention of N in the lunar mantle, and help it resist degassing and loss to space. 

\section{``Bottom-up'' approach: terrestrial analyses } \label{sec:bottomup}
 In this section, we present our ``bottom-up'' approach. That is, using analyses of geologic materials to estimate  the N budget of the Earth in its present state. We will not make a thorough attempt to describe how N cycling has changed in the past, but will discuss past cycles/characteristics where needed. We organize our budget starting with the best characterized reservoirs (atmosphere, ocean, biomass) then describing those that are less well known (geologic). Bulk reservoir masses (Table \ref{tab:assumptions}) are used in concert with estimated N concentrations to calculate N mass in poorly characterized reservoirs  (Table \ref{tab:total}). A more complete picture of the current state of N on Earth should provide a more solid springboard from which to leap into interpretations of past processes.  

\subsection{Atmosphere} \label{sec:atmosphere}

N$_2$ is the dominant form of N in the atmosphere; its mass ($4\times10^{18}$ kg) is calculated via Equation \ref{eq:atmosphere}: using 
 m$_{\textrm{a}}$ as  molar mass  dry air (0.02897 kg mol$^{-1}$); $x_{\textrm{N}_2}=0.78$; $\textrm{r}=6.4\times10^6$ m; $\textrm{p}=1.014\times10^5$ Pa; and $\textrm{g}=9.8~\textrm{m}~\textrm{s}^{-2}$.  

Other N species in the atmosphere include N$_2$O, NH$_Y$ (NH$_3$, NH$_4^+$), and NO$_X$ (NO, NO$_2^-$, NO$_3^-$). These are minor species, with abundances of 1.5$\times10^{12}$, 1.7$\times10^{9}$, and 7$\times10^{8}$ kg N, respectively \citep{Ussiri_and_Lal_2013}. 
\begin{table*}

\caption{Well characterized surficial N reservoirs, including the atmosphere, ocean, and biomass.}
\centering
\begin{tabular}[h]{ l  c c }  
\hline\\

\bf{Reservoir}	&	\bf{Form}	&	\bf{Size ($10^{18}$ kg N)}	\\[2ex] \hline 
Atmosphere	&	N$_2$	&	4.0	\\
	&	N$_2$O	&	1.5$\times10^{-6}$	\\
	&	NH$_Y$	&	1.7$\times10^{-9}$	\\
	&	NO$_X$	&	7$\times10^{-10}$	\\[1ex]
Oceans	&	N$_2$	&	2.4$\times10^{-2}$	\\
	&	NO$_3$	&	5.7$\times10^{-4}$	\\
	&	NH$_Y$	&	7$\times10^{-6}$	\\
	&	N$_2$O	&	2$\times10^{-7}$	\\
Biomass	&	Ocean Living	&	5$\times10^{-7}$	\\
	&	Ocean Dead Organic Matter	&	8$\times10^{-4}$	\\[1ex]
	&	Continental Living	&	1.1$\times10^{-5}$	\\
	&	Soil organics	&	1.3$\times10^{-4}$	\\[1ex]
\hline					
			
\label{tab:known}
\end{tabular}
\end{table*}

\subsection{Oceans}
The N content of  the oceans  is small compared to the atmosphere. 
Dissolved N$_2$ is the main N species in the ocean, with a mass of about 2.4$\times10^{16}$ kg  \citep{Ussiri_and_Lal_2013}. Concentrations of minor species (NO$_3^-$, NH$_4^+$, N$_2$O) can vary over the year, spatially, and with depth.  Surface NO$_3^-$ concentration is typically higher in the winter, due to lower productivity, but varies throughout the year at concentrations of 0-30 $\mu$M \citep{Garcia_et_al_2010}.  NO$_3^-$ at depth is more constant spatially and is found at higher concentrations, between 10-35 $\mu$M. Total NO$_3^-$ in the  ocean has a mass of 5.7$\times10^{14}$ kg N \citep{Ussiri_and_Lal_2013}.  On average, the NH$_4^+$ is found at a concentration of 0.4 $\mu$M, and N$_2$O at a concentration of 11 nM. These concentrations yield masses of  7$\times10^{12}$ and 2$\times10^{11}$ kg N, respectively. 

Concentrations of N species also depend on redox conditions. In deep waters of the Black Sea, which are anoxic, NH$_4^+$ is the dominant N species, with concentrations of up to 40 $\mu$M, while NO$_3^-$ concentration is negligible \citep{Fuchsman_et_al_2008}.  A similar relationship is seen during  the winter in Saanich Inlet on Vancouver Island, with deep anoxic waters dominated by NH$_4^+$ (10 $\mu$M) instead of NO$_3^-$ ($<0.2~\mu$M) \citep{Velinsky_et_al_1991}. 

\subsection{Biomass} 
The mass of N in living biomass is small compared with dissolved N$_2$ (above) and inorganic N species. Living biomass in the ocean contains about 5$\times10^{11}$ kg N. Marine dead organic matter is a much more substantial reservoir, comparable to inorganic fixed N, with 8$\times10^{14}$ kg N \citep{Ussiri_and_Lal_2013}. 

Soil and terrestrial biomass constitute a  reservoir of N comparable to oceanic biomass. Soil has a N content of 1.62$\times10^{14}$ kg  and living biomass has a N mass of 1.1$\times10^{13}$ kg   \citep{Ussiri_and_Lal_2013} .
 
 Despite its low mass in biomass, we emphasize the importance of biology in fixing N. This process is responsible for transferring N from the atmosphere into other reservoirs in the Earth. 
 
\begin{center}
\begin{table*}
\caption{Physical characteristics of geologic reservoirs used to calculate N mass.}
\centering
\begin{tabular} {l c c c c}
\hline \\
\bf{Reservoir} & \bf{Density (g/cm$^3$)} & \bf{Thickness (km)} & \bf{Area (km$^2$)} & \bf{Mass (kg)}  \\ [1ex]

Oceanic Sediments$^1$  &- &- &- & 7.4$\times10^{20}$  \\ [.5ex]
Oceanic Lithosphere & 3 & 50 &3.61$\times10^8$&  5.4$\times10^{22}$  \\  [.5ex]

Continental Crust$^2$ &- & -& -& $1.9\times10^{22}$  \\ [.5ex]

MORB-source Upper Mantle & 4 & 400 & $3.61\times10^8$ & $5.8\times10^{23}$  \\ [.5ex]
Off-cratonic Upper Mantle$^3$  & 4 & 400 & $6\times10^7$ & 9.6$\times10^{22}$  \\ [.5ex]
Cratonic Upper Mantle$^3$ & 4 & 400 &$9\times10^7$& 1.4$\times10^{23}$\\ [.5ex]
Transition Zone$^4$ & 4.4 & 240 & volume = $1.1\times10^{11}~\textrm{km}^3$  & 4.8$\times10^{23}$   \\[.5ex]
Lower Mantle &-&-&-&$2.93\times10^{24}$ \\[.5ex]
\hline
\multicolumn{5}{c}{\small References are: $^1$\cite{Veizer_and_Mackenzie_2003}, $^2$\cite{Taylor_and_McLennan_1995}, $^3$\citep{Lee_et_al_2011}, $^4$\cite{Gu_and_Dziewonski_2001}    } \\

\label{tab:assumptions}
\end{tabular}
\end{table*}
\end{center} 

\subsection{The Crust}
\subsubsection{Oceanic Sediments}

Nitrogen concentration in oceanic sediments is ultimately controlled by local biologic activity. In the modern ocean, primary productivity is higher near continental margins. Proximity to continental margins increases the available nutrient pool via river/weathering input and upwelling nutrient-rich deep waters \citep{Gruber_and_Sarmiento_1997}. Consequently, primary productivity is higher near continental margins. Higher productivity leads to  greater organic matter concentration, and the potential for more N to be deposited in sediments. Indeed, concentrations of N in sediments off the Central American margin have  $>$1000 ppm N in some locations, with an average of about 770 ppm \citep{Li_and_Bebout_2005}. This is notably higher than sediments from the Izu-Bonin-Mariana arc (IBM) in the western Pacific, which is further from continental margins, has lower primary productivity, and an average N content of about 280 ppm \citep{Sadofsky_and_Bebout_2004}. 

\begin{table*} 
\caption{Concentration of N in  oceanic sediments, crust, and lithospheric mantle. All are shown with standard error of the mean (SE$\bar{x}$).  Samples used in budget estimates are 250 Ma or younger, as this is the maximum age of oceanic crust. Oceanic crust and altered includes basalts and gabbros. Metamorphosed oceanic crust comprises basalts, gabbros, blueschists, and eclogites. MORB-source mantle includes peridotite (harzburgite or uncategorized), while altered MORB-source mantle also includes serpentinite. Note that metamorphosed oceanic crust and altered MORB-source mantle samples may be older than 250 Ma.} 
\centering
\begin{tabular}{l c c c } \hline \\
 &\multicolumn{2}{c}{ \bf{Concentration (ppm)} } &\\
\bf{Rock Type} & \bf{$\bar{x}$}  & \bf{SE$\bar{x}$} & \bf{No. samples } \\ \hline \\ 
Biogenic & 1930 & 1540 & 161$^{1-3}$\\
Clastics & 570 & 56 &82$^{4-6}$ \\
Carbonates & 130 & 17 & 21$^{6-8}$ \\ 
\emph{Sediment Average} &560&230&\\[1ex]
Oceanic Crust &  1.4&1.3&124$^{9-17}$\\
Altered Oceanic Crust & 6.1 &0.7 &63$^{\textrm{16,18,19}}$\\
Metamorphosed Oceanic Crust & 7.1 &1.2 &14$^{\textrm{17,18,20,21}}$\\
 Oceanic Lithospheric Mantle & 0.24&0.33 &10$^{\textrm{12,21}}$\\
Altered Oceanic Lithospheric Mantle & 3.7 &0.5 &17$^{\textrm{17,18,20,22}}$\\
\hline

\multicolumn{4}{c}{\small{References: $^1$\cite{Peters_et_al_1978}, $^2$\cite{Chicarelli_et_al_1993}, $^3$\cite{Quan_et_al_2008}, $^4$\cite{Sullivan_et_al_1979}, }}\\
\multicolumn{4}{c}{\small{$^5$\cite{Sadofsky_and_Bebout_2004}, $^6$\cite{Li_and_Bebout_2005}, $^7$\cite{Rau_et_al_1987}, $^8$\cite{Sephton_et_al_2002}}}\\
\multicolumn{4}{c}{\small{$^9$\cite{Sakai_et_al_1984}, $^{10}$\cite{Exley_et_al_1987}, $^{11}$\cite{Marty_1995}, $^{12}$\cite{Marty_and_Humbert_1997}}}\\
\multicolumn{4}{c}{\small{$^{13}$\cite{Sano_et_al_1998}, $^{14}$\cite{Marty_and_Zimmerman_1999}, $^{15}$\cite{Nishio_et_al_1999}, 
$^{16}$\cite{Li_et_al_2007},  }}\\
\multicolumn{4}{c}{\small{$^{17}$\cite{Halama_et_al_2012}, $^{18}$\cite{Busigny_et_al_2003}, $^{19}$\cite{Busigny_et_al_2005a}, $^{20}$\cite{Halama_et_al_2010},}}\\
\multicolumn{4}{c}{ \small{$^{21}$\cite{Busigny_et_al_2011}, $^{22}$\cite{Philippot_et_al_2007}  }}\\
\label{tab:occrust}
\end{tabular}
\end{table*} 

\begin{figure*}
\centering
\includegraphics[keepaspectratio=true, height=3.2in]{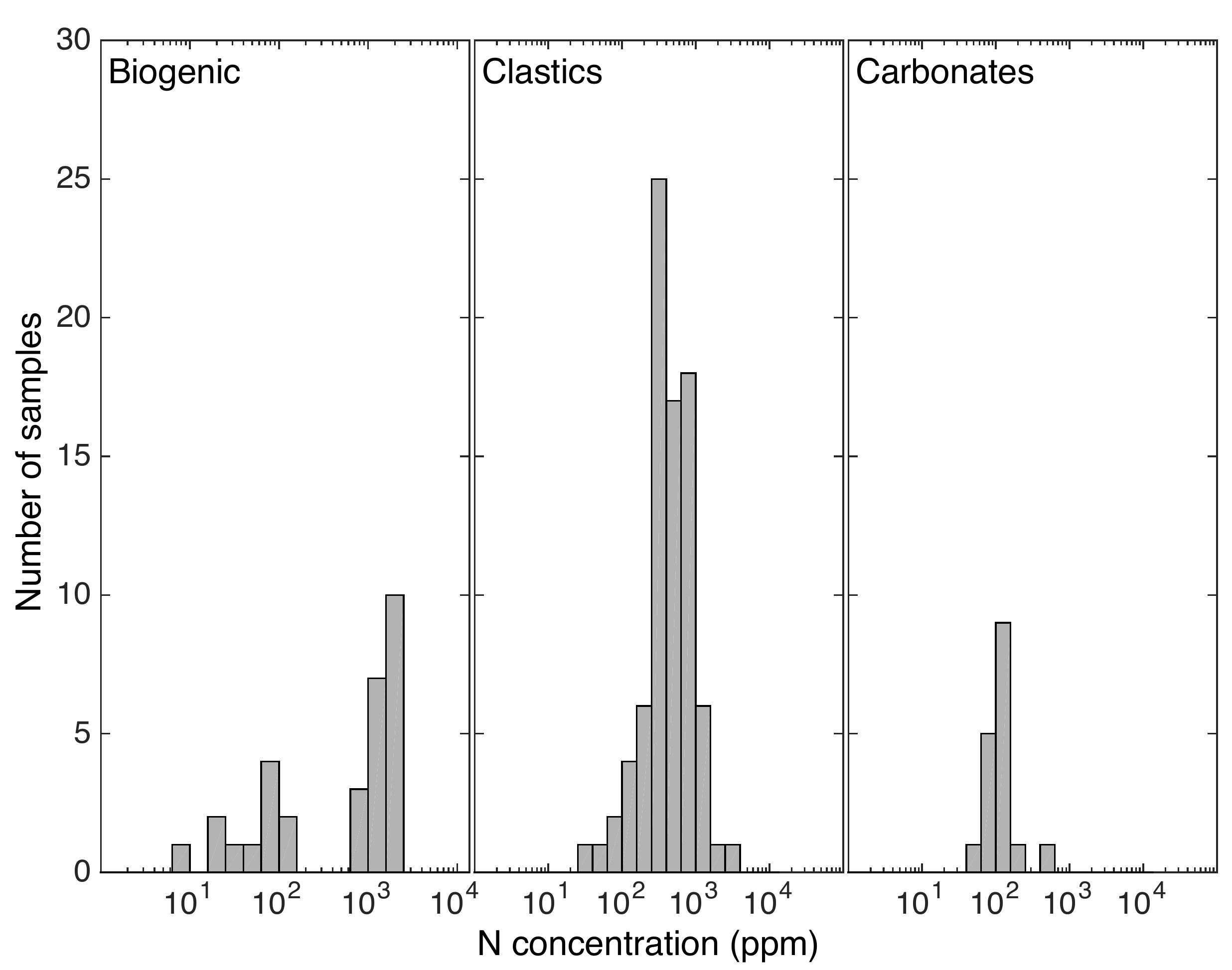}
\caption{Nitrogen concentrations in oceanic crust less than 250 Ma. Sediments are separated into biogenic (organic materials, kerogen, graphite, microbialite, chert), clastic (siltstone, mudstone, clay, shale, sandstone, and pelite), and carbonates (limestone, dolostone). Oceanic lithosphere samples are labeled as OL and AOL. }
\label{fig:ocseds}
\end{figure*}

Our N abundance estimate in oceanic sediments is based on the proportion of three sediment types covering the sea floor.  All samples that are younger than 250 Ma are used, since this is the maximum age of oceanic basins. Current sedimentary cover on the sea floor  is composed of carbonate (47.1$\%$), clastic (38.1$\%$), and biogenic ($14.8\%$) sediments \citep{Davies_and_Gorsline_1976, Brown_et_al_1989}.  Biogenic sediments include organic materials, kerogen, graphite, microbialite, and chert and  have a mean N concentration of 1930$\pm1540$ ppm (Fig \ref{fig:ocseds}, Table \ref{tab:occrust}). Clastic sediments include siltstone, mudstone, clay, shale, sandstone, and pelite and have a mean N concentration of 570$\pm56$ ppm. Carbonates, limestone and dolostone, have a mean N concentration of 130$\pm17$ ppm N. The resulting weighted average is 560$\pm230$ ppm, which yields a N content for oceanic sediments of 0.41$\pm0.2\times10^{18}$ kg (Table \ref{tab:total}).  

\begin{figure*}
\centering
\includegraphics[keepaspectratio=true, height=3.2in]{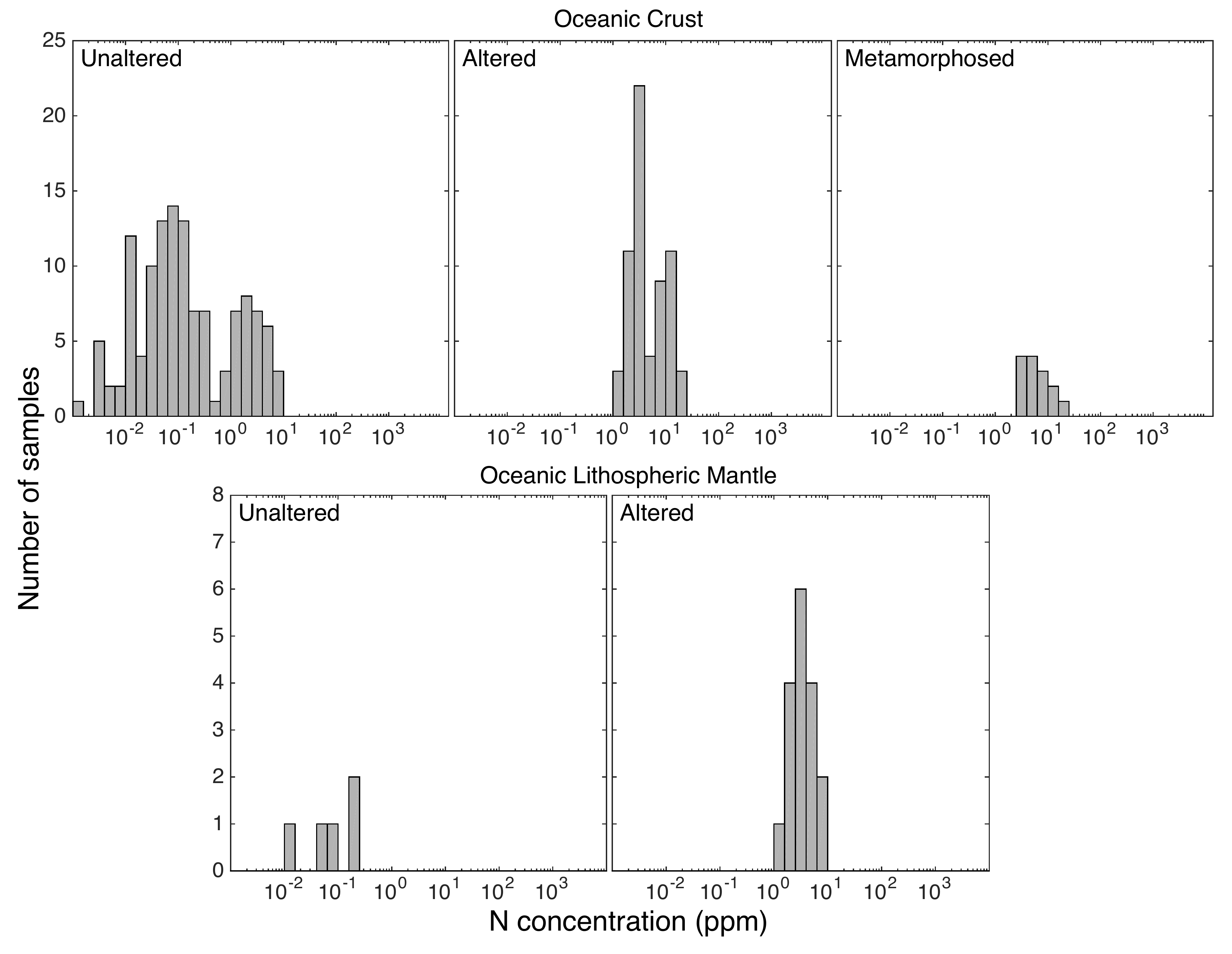}
\caption{Nitrogen concentrations in oceanic crust and lithospheric mantle. Hydrothermal alteration increases the N content of oceanic crust, and it retains high N even during metamorphism. The same increase is seen in the oceanic lithospheric mantle. Samples are as in Table \ref{tab:occrust}. }
\label{fig:occrust}
\end{figure*}

An important consideration when attempting to estimate N contents in the past is the redox conditions of the overlying water column, as sediments deposited under anoxic water conditions may be more N-rich. A sediment  core from the Black Sea shows that sediments deposited under an anoxic water column have twice the N content (1800 ppm) as  sediments deposited under an oxic water column (900 ppm) \citep{Quan_et_al_2013a}.  Changes in redox sensitive metal (Fe, Mo) concentrations and lack of significant changes in total organic carbon and $\delta^{13}$C values corroborate redox control over N concentration, as opposed to a purely biological control. The increase of N content with reducing conditions may not be universal, as some  shale units  do not show N enrichment under anoxic conditions. They do, however, show distinct $\delta^{15}$N values that appear to reflect redox conditions \citep{Quan_et_al_2013b}.

\subsubsection{Oceanic Crust and Upper Lithosphere} \label{sec:occrust}

Fresh gabbros and basalts at mid-ocean ridges  inherit N from their mantle source \citep{Marty_1995}. Since these magmas are oxidized,  N present in fresh MORB is typically found as N$_2$ locked in fluid inclusions in glassy rims. Concentrations in fresh crust are low, with an average of $1.4\pm1.3$ ppm. Concentration is unaltered lithospheric mantle (defined here as harzburgite and undifferentiated peridotite) is also low, at $0.24\pm0.33$ ppm (Fig. \ref{fig:occrust}, Table \ref{tab:occrust}).

Hydrothermal alteration tends to increase the N content of the rocks, to an average of $6.1\pm0.7$ ppm for crustal rocks (altered basalts and gabbros) and $3.7\pm0.5$ ppm for lithospheric mantle (harzburgite and serpentinite). The source of this N is from biologic activity in seawater,  identified by  positive $\delta^{15}$N values. As seawater enters a hydrothermal system, it carries NH$_4^+$ from overlying sediments into the crust and mantle \citep[e.g.,][]{Halama_et_al_2010}. NH$_4^+$ substitutes into mineral lattices of hydrothermal minerals, most importantly K-bearing clays. It is possible that some N may be present as N$_2$ in cordierite \citep{Palya_et_al_2011}. 

 Since hydrothermal alteration is the main control on geologically stable N in oceanic crust, estimates of  N concentration in these rocks  depend  on the depth and extent of hydrothermal alteration into the lithosphere. The deepest cores show that alteration occurs at least to a depth of 470 m \citep{Li_et_al_2007}. Metagabbros and serpentinized mantle rocks in ophiolites show that alteration can reach even deeper, into the upper mantle. Hydrothermal origin of N is confirmed by enriched $\delta^{15}$N values, derived from oceanic biologic processes \citep{Busigny_et_al_2011, Halama_et_al_2012}.  Some ophiolites experienced metamorphic pressures of up to 2.5 GPa ($\sim80-90$ km depth), yet  still retain N derived from the ocean, indicating the durability of the NH$_4^+$-silicate bond. Indeed, the concentration of N in metamorphosed oceanic crust (basalt, gabbro, blueschist, eclogite) is  identical within error ($7.1\pm1.2$ ppm) to altered crust.

  A N budget estimate for the oceanic lithosphere can be calculated assuming shallow or deep alteration. Shallow alteration affecting the entire crust ($6.1\pm0.7$ ppm),  0.5 km of mantle ($3.7\pm0.5$ ppm),  with the remainder of the mantle (9.5 km) at $0.24\pm0.33$ ppm N, yields a N concentration of $2.9\pm0.3 $ ppm. This concentration gives N mass of  $0.16\pm0.02\times10^{18}$ kg N. If alteration occurs on a lithospheric scale (8 km crust and 10 km mantle), we calculate an upper estimate of $4.8\pm0.4$ ppm N, which gives total N mass of $0.26\pm0.02\times10^{18}$ kg N. While these values are orders of magnitude less than the amount of N contained in the atmosphere, N in the oceanic crust is of critical importance as  subduction  over long time scales has the potential to transport a large amount of N into the mantle.  
  
 An average column of oceanic crust, sediments, and lithosphere can subduct substantial N over Earth history. A column with 500 m of sediment (560 ppm N, Table \ref{tab:occrust}) and the conservative oceanic crust plus lithosphere concentration  of $2.9\pm0.3$ ppm gives an average column concentration of 18 ppm. If we assume that all N gets subducted, which depends on temperature and varies by subduction zone, we can multiply this concentration times the mass of oceanic slab being subduction each year ($\sim40,000$ km trench length, 18.5 km thick crust, 5 cm/yr convergence rate, and density of 3.5 g/cm$^3$), and calculate that $2.3\times10^{9}$ kg N are subducted every year. Over Earth history (4 Ga, for illustrative purposes), current subduction equates to $9.3\times10^{18}$ kg N, which is twice the current mass of N in the atmosphere. It therefore seems reasonable to suggest that the entire atmosphere may have passed through the mantle at least once, given current subduction efficiency, or more times if subduction of N was more efficient in the past (Sec. \ref{sec:discussion}).

\subsubsection{Continental Crust}

The continental crust is composed of two  categories of rocks: (meta)sedimentary  and (meta)igneous. We base our estimate for the N budget of the continental crust on our literature compilation and the rock abundance estimates of \cite{Wedepohl_1995} (Table \ref{tab:contcrust}). These proportions are based on surface outcrop area for upper crustal rocks and xenolith data for lower crustal rocks. 

\begin{figure*}

\centering
\includegraphics[keepaspectratio=true, width=\textwidth]{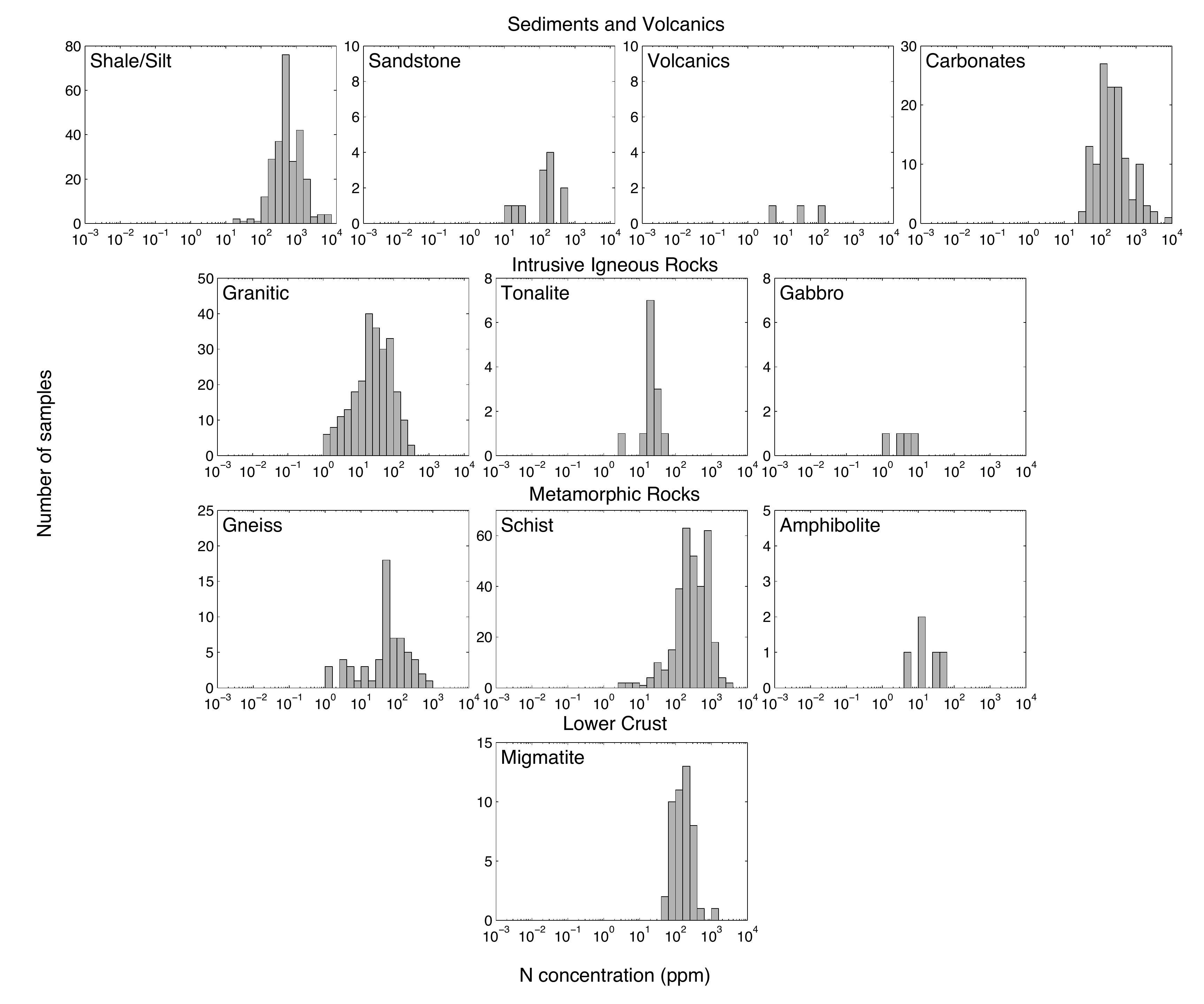}
\caption{N concentrations in continental crust. Panels are organized from shallow to deep levels in the crust. See Table \ref{tab:contcrust} for average N concentrations.  }
\label{fig:contcrust}

\end{figure*}

\newgeometry{left=0.75in,right=0.75in,top=1in,bottom=1in}
\begin{table*}

\centering
\small
\caption{ Estimates for the amount of N in the continental crust, shown with standard error of the mean (with an arbitrary error of 50$\%$ for very poorly known rock types). Rock proportions are based on \cite{Wedepohl_1995}, who based upper crustal rock abundances on mapped area and lower crustal abundance on xenoliths.  }

\begin{tabular}{l c c c l } \hline \\

&&\multicolumn{2}{c}{\bf{Conentration (ppm)}}&\\
\bf{Reservoir ($\%$ of crust)} & \bf{Rock type ($\%$ of reservoir)} & \bf{$\bar{x}$}  & \bf{SE$\bar{x}$} & \bf{No. samples} \\ \hline \\ 
\bf{Upper Crust ($53\%$)} & & &  \\ [1ex]
Sedimentary/Volcanic Rocks (14$\%$) & Shale/Silt (44$\%$) & 860  & 64& 261$^{1-11}$\\ 
                                                & Sandstone (21$\%$) & 230 & 110 & 12$^{12-16}$\\  
                                                & Volcanics ($20\%$) & 50 &  60 & 3$^{2,14}$  \\
                                                       & Carbonates (14$\%$) & 130   & 17& 17$^{\textrm{see Table \ref{tab:occrust}}}$ \\ 
                                                & \emph{Average} & 455 & 38&\\ [1ex]
  Felsic Intrusives (50$\%$) & Granitic (90$\%$) & 54  & 7 & 247$^{14,17-29}$\\
  					  & Tonalite (10$\%$) & 24 &   4 &13$^{2,18,27-28}$ \\
					  & \emph{Average} & 52 &6&  \\ [1ex]
Mafic Intrusives (6$\%$) & Gabbros (100$\%$) & 5 &  2 & 4$^{2,28,30-31}$  \\[.5ex]
Metamorphic rocks (30$\%$) & Gneisses (64$\%$) & 135 & 50& 63$^{2,17-18,32-37}$ \\
					     & Mica Schist (16$\%$) & 500& 44  & 323$^{2,14,17-18,28,31-33,35-37,38-48}$ \\
					     & Amphibolites (18$\%$) & 22 & 10& 5$^{17,42,49}$ \\
					     & Marble (3$\%$) & 1000 &  500(?) & 1$^{50}$ \\
					     & \emph{Average} & 200 & 36& \\[1ex]
\bf{Upper Crust Average} & & 150  & 12& \\[1ex]
 
 \bf{Lower Crust ($47\%$)} & & &&  \\[1ex] 
 Felsic/Mafic Granulites (62$\%$) &  &   17  &6&   22$^{22,27,32,37,47,51-56}$\\
 Mafic Granulites ($38\%$) & & 17&6& 22$^{22,27,32,37,47,51-55,57}$\\[1ex]
 \bf{Lower Crust Average} & & 17 &6& \\[1ex]

 \bf{Total crust average} & & 120 &9&  \\
& \multicolumn{3}{l}{\bf{Continental Crust N estimate$\mathbf{=1.7\pm0.1\times10^{18}}$ kg N}}& \\

\hline

\multicolumn{5}{l}{\tiny References are: $^1$\cite{Itihara_et_al_1986}, $^2$\cite{Honma_1996}, $^3$\cite{Watanabe_et_al_1997},  $^4$\cite{Yamaguchi_2002}, $^5$\cite{Krooss_et_al_2005}, $^6$\cite{Kerrich_et_al_2006}, $^7$\cite{Garvin_et_al_2009}}\\
\multicolumn{5}{l}{\tiny$^8$\cite{Papineau_et_al_2009}, $^9$\cite{Busigny_et_al_2013}, $^{10}$\cite{Cremonese_et_al_2013},$^{11}$\cite{Godfrey_et_al_2013} $^{12}$\cite{Sullivan_et_al_1979}, $^{13}$\cite{Greenfield_1988}, $^{14}$\cite{Greenfield_1991},}\\
\multicolumn{5}{l}{\tiny $^{15}$\cite{Williams_et_al_1995}, $^{16}$\cite{Pontes_et_al_2009}, $^{17}$\cite{Itihara_and_Honma_1979}, $^{18}$\cite{Honma_and_Itihara_1981}, $^{19}$\cite{Hall_1987}, $^{20}$\cite{Cooper_and_Bradley_1990}, $^{21}$\cite{Hall_et_al_1991} }\\

\multicolumn{5}{l}{\tiny$^{22}$\cite{Boyd_et_al_1993}, $^{23}$\cite{Hall_et_al_1996}, $^{24}$\cite{Bebout_et_al_1999a}, $^{25}$\cite{Hall_1999},$^{26}$\cite{Jia_and_Kerrich_1999},$^{27}$\cite{Jia_and_Kerrich_2000}, $^{28}$ \cite{Ahadnejad_et_al_2011},}\\
\multicolumn{5}{l}{\tiny $^{29}$\cite{Morford_et_al_2011}, $^{30}$\cite{Halama_et_al_2010}, $^{31}$\cite{Busigny_et_al_2005b}, $^{32}$\cite{Duit_et_al_1986}, $^{33}$\cite{Haendel_et_al_1986}, $^{34}$\cite{Visser_1993}, $^{35}$\cite{Mingram_and_Brauer_2001},}\\
\multicolumn{5}{l}{\tiny $^{36}$\cite{Jia_2006}, $^{37}$\cite{Cruz_2011}, $^{38}$\cite{Bebout_and_Fogel_1992}, $^{39}$\cite{Bebout_1997}, $^{40}$\cite{Boyd_and_Philippot_1998}, $^{41}$\cite{Bebout_et_al_1999b}, $^{42}$\cite{Holloway_et_al_2001},}\\
\multicolumn{5}{l}{\tiny $^{43}$\cite{Busigny_et_al_2003}, $^{44}$\cite{Sadofsky_and_Bebout_2003}, $^{45}$\cite{Pitcairn_et_al_2005}, $^{46}$\cite{Yui_et_al_2009}, $^{47}$\cite{Plessen_et_al_2010}, $^{48}$\cite{Palya_et_al_2011}, $^{49}$\cite{Busigny_et_al_2011} }\\

\multicolumn{5}{l}{\tiny$^{50}$\cite{Dixon_et_al_2012}, $^{51}$\cite{Itihara_and_Tainosho_1989}, $^{52}$\cite{Hall_1999}, $^{53}$\cite{Sadofsky_and_Bebout_2000}, $^{54}$\cite{Pinti_et_al_2001}, $^{55}$\cite{Li_et_al_2014},}\\
\multicolumn{5}{l}{\tiny For mineral modes: $^{56}$\cite{Galli_et_al_2011}, $^{57}$\cite{Hansen_and_Stuk_1993}}\\

\label{tab:contcrust}
\end{tabular}
\end{table*}
\restoregeometry

Upper crustal rocks are, unsurprisingly, the most analyzed and well characterized. Clastic sediments (especially shales) are the most sampled, and these include both samples explicitly formed in continental settings and sediments that formed in oceanic settings and are older than 250 Ma. Samples with oceanic provenance older than 250 Ma must now be preserved on continents in order to be sampled, so are included in the continental N budget. Nitrogen in continental sedimentary rocks is  incorporated as organic matter, NH$_4^+$ from the breakdown of organic matter (as described previously), or NH$_4^+$ in minerals weathered from crystalline rocks. 

Crystalline rocks, both igneous and metamorphic, form the majority of the upper continental crust. Nitrogen in both rock types is predominantly NH$_4^+$, either inherited from source rocks (for igneous) or protoliths (for metamorphic). The presence of biotite exerts a strong control over the distribution of N, as biotite has a strong affinity for NH$_4^+$ when compared to other K$^+$-bearing minerals \citep{Honma_and_Itihara_1981}. Retention of NH$_4^+$  in a source region is also promoted by reduced melting conditions. In contrast, more oxidized melt conditions or less biotite-rich sources should preferentially move NH$_4^+$ into the melt, and therefore into granites, though some NH$_4^+$  may oxidize to N$_2$, and either escape the melt or become trapped in fluid inclusions \citep{Hall_1999}. 

Hydrothermal alteration, especially if fluids pass through sedimentary country rock, will increase the concentration of N in crystalline rocks. For example, globally, unaltered granitic rocks  have an  average N concentration of $36\pm4$ ppm; globally, altered granitic rocks average $65\pm17$ ppm N, with concentrations up to 250 ppm \citep[e.g.,][]{Boyd_et_al_1993, Holloway_and_Dahlgren_2002}.

N analyses from the deeper crust are sparse, but recent measurements suggest that this might be an important long-term reservoir for N \citep{Palya_et_al_2011}.  
To calculate the mass of lower crust N, we use two well studied outcrops to represent the average mineralogy of both mafic (Cone Peak, California) \citep{Hansen_and_Stuk_1993} and felsic (Gruf Complex, eastern Alps) \citep{Galli_et_al_2011} granulites. The majority of N in these rocks should be contained in biotite and potassium feldspar, with some in plagioclase in a ratio of 1:0.38:0.11 \citep{Honma_and_Itihara_1981}.  Mafic granulites have 17$\%$ biotite, $59\%$ plagioclase, and negligible potassium feldspar; felsic granulites have $17\%$ biotite, $17\%$ plagioclase, and $10\%$ potassium feldspar. Globally, average N content of biotite in metamorphosed continental igneous and sedimentary rocks is $87\pm22$ ppm, which in turn suggests plagioclase (in equilibrium) has $10\pm2$ ppm and potassium feldspar $33\pm8$ ppm. Thus, mafic and felsic granulites have similar N concentrations of $17\pm6$  ppm (Table \ref{tab:contcrust}).

Our total continental crust estimate, $1.7\pm0.1\times10^{18}$ kg N (Table \ref{tab:contcrust}), suggests a substantial amount of N may be sequestered in the continental crust. This estimate is equivalent to a recent rough estimate \citep{Goldblatt_et_al_2009}, but higher than another recent study \citep{Rudnick_and_Gao_2014} based on older compilations \citep{Wedepohl_1995} and measurements \citep{Wlotzka_1972}. 
				
\subsection{The Mantle}\label{sec:TheMantle}
The large mass of the mantle, compared to the atmosphere, means that it contains substantial N, even
at low concentration. For example, 1 ppm N in the mantle would give N mass of $4\times10^{18}$ kg, which is the same as the atmosphere. Determining the actual abundance of N in the mantle is difficult,  as N analyses are rare. Most information concerning volatile elements is from noble gas geochemistry, which is augmented by  diamond analyses, xenoliths, and direct mantle melts. We can estimate the N content of the mantle either as a whole or by  breaking it into different domains. First we calculate a whole-mantle estimated based on N-Ar geochemistry. Then, we discuss distinct domains (MORB-source, OIB-source, off-cratonic, cratonic) individually, and base N estimates on analyses of xenoliths where available, as well as detail potential N capacity in unsampled domains (transition zone, lower mantle).

\begin{figure*}
\centering
\includegraphics[keepaspectratio=true, width=0.75\textwidth]{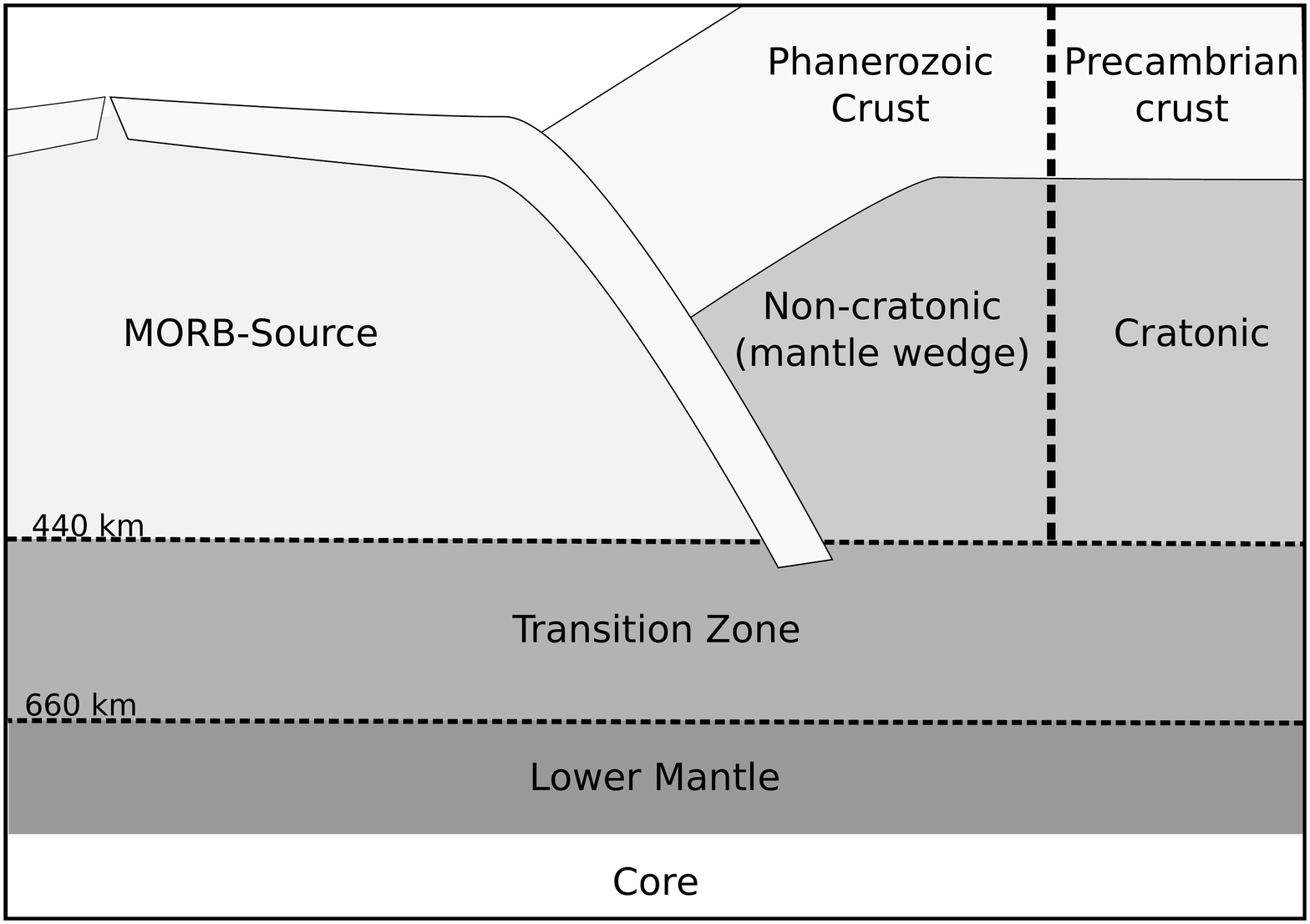}
\caption{Mantle reservoirs as defined for individual domain-based budget (Section \ref{sec:indv}). The transition zone and lower mantle are defined by mineral phase changes. The mass of non-cratonic and cratonic mantle is estimated by multiplying the area of Phanerozoic-aged crust and pre-Cambrian crust by the depth to the top of the transition zone.}
\label{fig:mantlereservoirs}
\end{figure*}

\subsubsection{Argon geochemistry-based estimate}\label{sec:NAr}

 Nitrogen and Ar behave similarly  in basaltic melts under oxidizing conditions \citep[e.g.,][]{Libourel_et_al_2003, Roskosz_et_al_2006, Mysen_and_Fogel_2010,  Li_et_al_2014, Li_et_al_2015}, so N-Ar systematics may be used to calculate whole-mantle N budgets \citep{Marty_1995, Marty_and_Humbert_1997, Dauphas_and_Marty_1999, Marty_and_Zimmerman_1999, Goldblatt_et_al_2009}. This approach is valid for both MORB and OIB, since they are generated from melting of oxidized upper mantle, even though they are geochemically distinct  \citep{White_2010}. There are a number of measurements of N$_2$ and Ar in both basalt types, which can be used to estimate N concentration in their source. Calculating the amount of N in MORB-source and OIB-source mantle depends on establishing three criteria: (a) the amount of Ar in MORB- and OIB-source mantle, (b) the relationship between N and Ar in MORB and OIB,  and (c) the proportion of mantle that is MORB- and OIB-source. 

Before calculating the N mantle budget, we highlight some aspects of Ar geochemistry, as these are crucial to the following interpretation. Argon has three isotopes, $^{36}$Ar, $^{38}$Ar, and $^{40}$Ar. The first two are primordial (i.e., inherited during planetary formation), while the third  is produced by radioactive decay of $^{40}$K, and has been increasing over time.  Both primordial isotopes are found almost exclusively in the atmosphere, though minor amounts are degassing from the mantle. The radiogenic isotope, $^{40}$Ar, is present in both the atmosphere and the BSE, since K is found in the solid Earth. Thus, we chose to compare N to $^{40}$Ar, as both elements are found in the atmosphere and BSE; these data are normalized to primordial $^{36}$Ar.

The first step needed to estimate N content from $^{40}$Ar is to calculate the amount of $^{40}$Ar present in the mantle. As mentioned, all $^{40}$Ar has been produced from the decay of $^{40}$K over time. Based on U/K ratios, the K concentration of the BSE is estimated at $280\pm120$ ppm \citep{Arevalo_et_al_2009}.  Given a known decay rate of $^{40}$K ($\lambda=5.55\times10^{-10}~\textrm{yr}^{-1}$), the proportion of this decay which produces $^{40}$Ar ($10.72\%$), and the abundance of $^{40}$K ($0.0117\%$) \citep{CRC_2014}, we  calculate  that a total of $4.2\pm1.8\times10^{18}$ mol $^{40}$Ar has been created over Earth history. Subtracting $^{40}$Ar in the atmosphere ($1.65\times10^{18}$ mol) and continental crust ($0.35\times10^{18}$ mol, \cite{Arevalo_et_al_2009}) gives the $^{40}$Ar content of the mantle to be $2.2\pm1.8\times10^{18}$ mol. 

\begin{figure*}[t]
\centering
\includegraphics[keepaspectratio=true, height=4.5in]{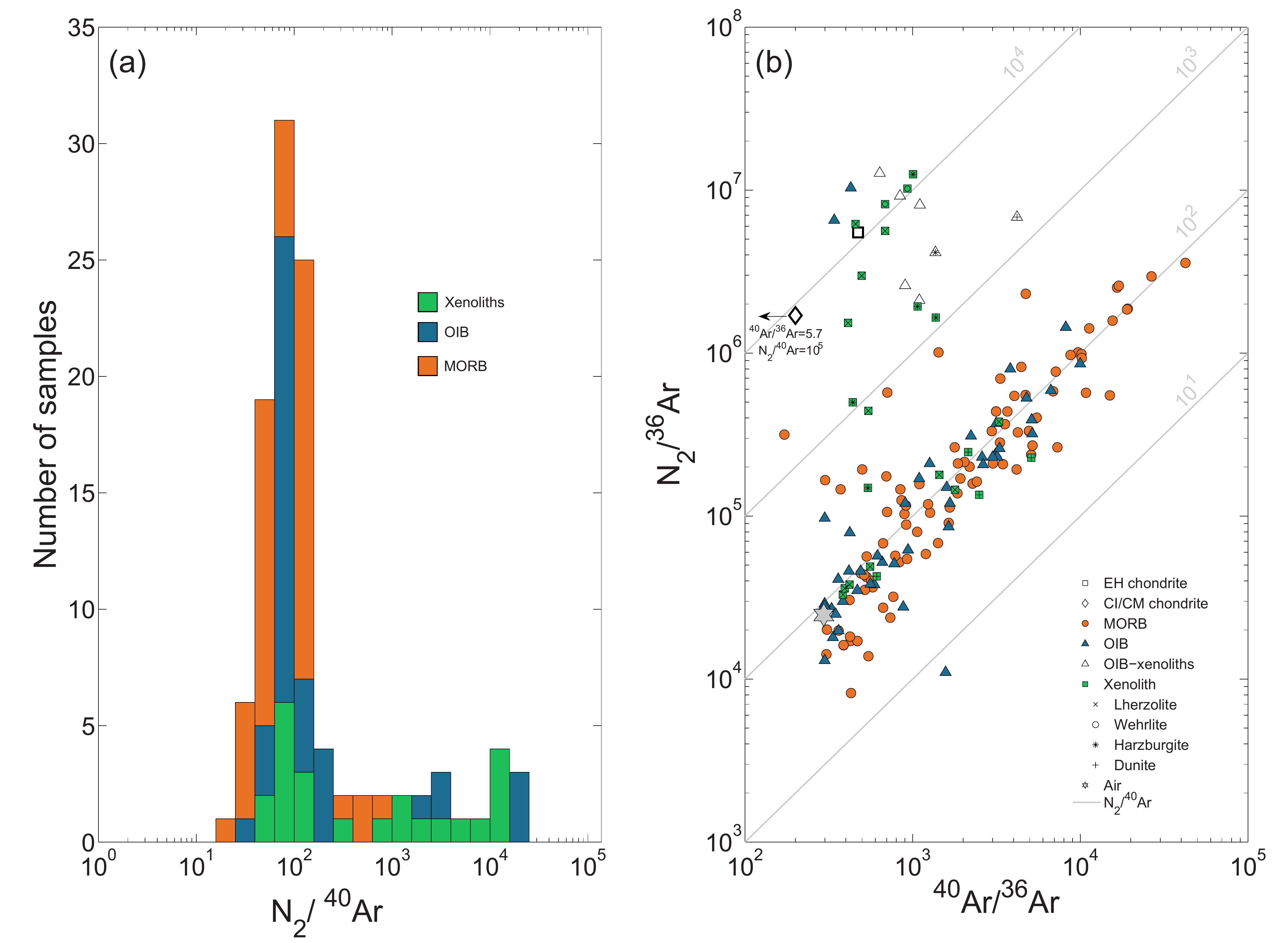}
\caption{(a) N$_2$ and Ar data showing log-normal distribution of N$_2$/$^{40}$Ar  MORB (orange),  OIB (blue), and xenolith (green) samples and (b) N$_2$/$^{36}$Ar as a function of $^{40}$Ar/$^{36}$Ar for the same rocks. MORBs are shown in orange circles, OIB basalts in blue triangles, OIB-associated xenoliths in white triangles, and other xenoliths in green squares. Additional symbols indicate known xenolith rock type. MORB samples show clear correlation, confirming previous results \citep{Marty_1995, Marty_and_Zimmerman_1999}. OIB samples both overlap  with MORB and define a distinct end-member. The existence of this nigh-N$_2/^{40}$Ar end member is corroborated by fertile xenolith samples. See text for discussion on the origin of this reservoir.}
\label{fig:N2Ar}
\end{figure*}

Next, we observe that the N$_2$ and $^{40}$Ar data fall into two populations (Fig. \ref{fig:N2Ar}): one containing MORB, some OIB, and some xenoliths with N$_2/^{40}$Ar values around $10^2$, which we coin as MORB-source like (MSL); and one containing some OIB and xenoliths with N$_2/^{40}$Ar around $10^4$, which we coin high-N. Interestingly, the  MORB samples fall along a coherent trend with air at one end, as seen in previous work \citep{Marty_1995}. Perhaps this indicates that the atmospheric N and Ar composition is the result of degassing the MORB-source mantle over time. In addition, the correlation between N$_2$ and $^{40}$Ar in the MORB, with weak correlation between N$_2$ and $^{36}$Ar  over the same range, indicates that N has been cycled through the mantle; by proxy, it correlates with K$^+$, which is concentrated in the continental crust, so observing a signal of K-input suggests contribution of continental material \citep{Marty_1995}. 

In contrast, OIB and xenolith data appear to describe a three-component mixture between air, a high-MSL end-member, and an end-member composition with high N$_2$ compared to $^{40}$Ar (Fig. \ref{fig:N2Ar}). Although the high-N field has only two OIB samples, we suggest it is a robust feature of the mantle as it is also defined by OIB-associated xenoliths and other xenoliths. OIB-associated xenoliths  are thought to represent OIB-source material on the basis of high $^3$He/$^4$He \citep[][ and references therein]{Mohapatra_et_al_2009}.  High-N xenoltih samples are  metasomatized, more fertile mantle lithologies (i.e., lherzolite, harzburgite, wehrlite) from locations in Oman and Europe \citep{Yokochi_et_al_2009}. In addition, MORB samples that fall off the MSL trend are analyses from the East Pacific Rise, which is thought to have a plume-like component \citep{Marty_and_Zimmerman_1999}, so may represent an intermediate between MSL and high-N mantle types.

To actually estimate N content in mantle source regions, we must determine the N$_2$/$^{40}$Ar ratio for MSL and high-N mantle (Fig. \ref{fig:N2Ar}). This is straightforward for MSL, which is described by a well-defined trend, and has a value of $120\pm11$. This value is consistent with previous work \citep[N$_2/\textrm{Ar}=124\pm6$, from][]{Marty_and_Zimmerman_1999}. Determining the high-N ratio is less straightforward, as it defines a more dispersed field. Since OIBs tend to record a somewhat more diverse set of mantle source types \citep{White_2010}, we suggest that taking the average N$_2/^{40}$Ar ratio from all samples with N$_2/^{40}$Ar$>10^3$ is the most appropriate approach to obtain a representative value. This ratio is $9.3\pm3.3\times10^3$.

\begin{landscape}
 \begin{figure*}[t]
\centering
\includegraphics[keepaspectratio=true, width=1.5\textwidth]{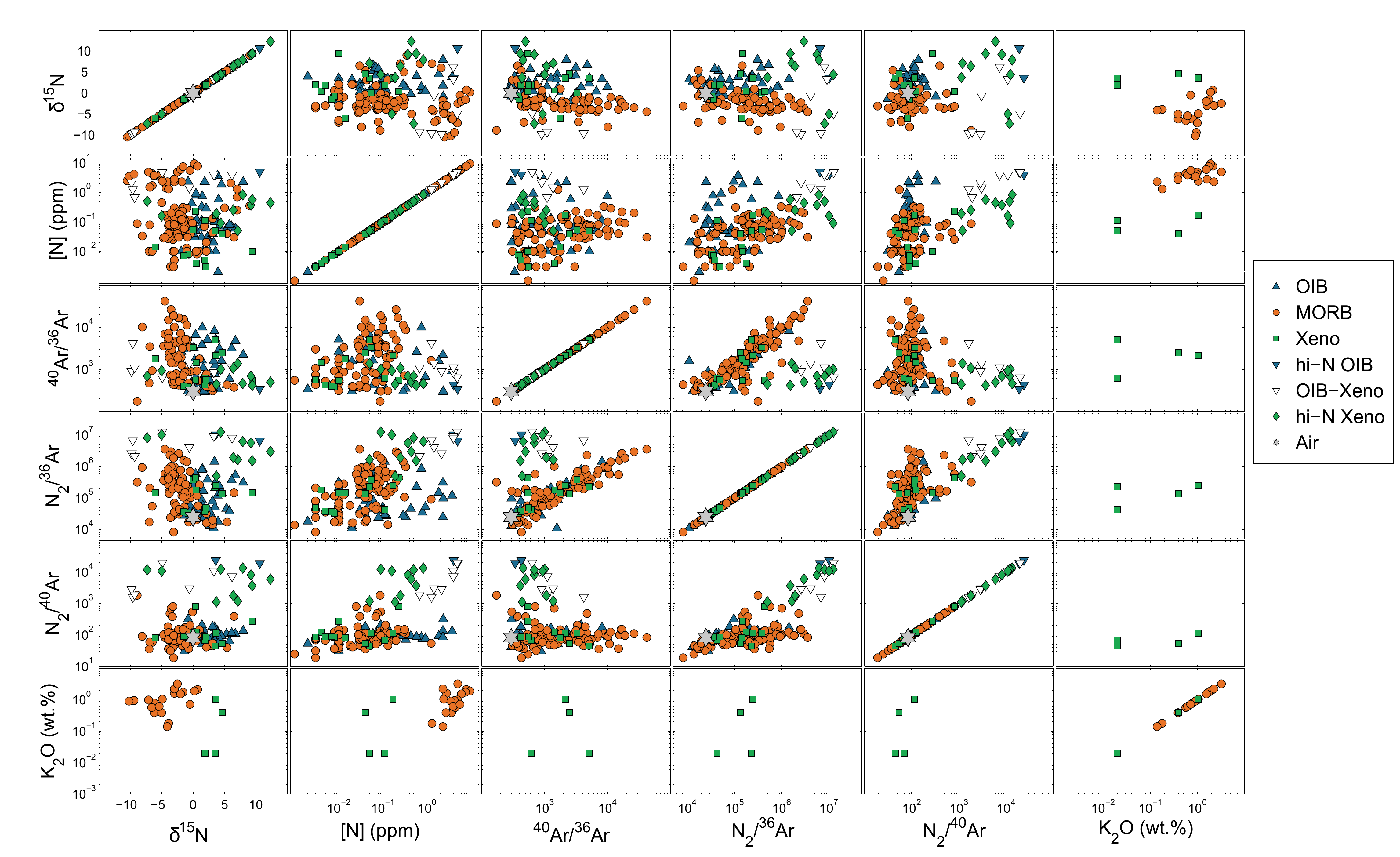}
\caption{All available N concentration, isotope, Ar-isotope, and K concentration data for MORB, OIB, and xenoliths. Two mantle N reservoirs identified in Fig. \ref{fig:N2Ar}: MORB-source like (MSL) and high-N are shown to separate in other plots as well. The high-N mantle tends to have high N concentration, low N$_2/^{40}$Ar, and variable but mostly positive $\delta^{15}$N values. We interpret these to mean this reservoir is young, N-rich, and derived from subducted surface materials. MSL mantle, by contrast, appears older, as indicated by low N$_2/^{40}$Ar. It is also primarily composed of depleted $\delta^{15}$N, which results from either incorporation of different subducted material \citep{Marty_and_Dauphas_2003} or different processing in the mantle }
\label{fig:N2ArFull}
\end{figure*}
\end{landscape}

 \begin{table*}[t]
\small
\caption{Nitrogen and Ar isotope data for CC and EC, as shown in Fig. \ref{fig:N2Ar}. CC data are from \citet{Marty_2012} and EC data from \citet{Hopp_et_al_2014}. We note that EC values are similar to our high-N mantle end member, while CC are not. This may be a coincidence, as N concentrations and $\delta^{15}$N values of high-N mantle and EC are not equivalent.  }
\centering
\begin{tabular}{l c c c c} \hline \\
\bf{Chondrite} & \bf{\ce{N2} (mol g$^{-1}$)}& \bf{\ce{N2}/$^{36}$Ar} & \bf{$^{40}$Ar/$^{36}$Ar} & \bf{\ce{N2}/$^{40}$Ar}\\
EC & $2.52\pm0.2\times10^{-5}$ & $5.5\pm0.4\times10^6$ & $473\pm10$ & $1.2\pm0.1\times10^4$  \\
CC & $4.4\pm0.1\times10^{-5}$ & $1.7\pm0.1\times10^6$ & $5.7\pm3.5$ & $3.1\pm1.3\times10^5$ \\ \hline

\end{tabular}
\end{table*}

Now armed with $^{40}$Ar abundance for the total mantle and N$_2/^{40}$Ar for MSL and high-N reservoirs, the last step required  is to estimate the actual proportion of these types of mantle. This is the most difficult of the three criteria. Trace element (e.g., U, Ta) mass balance suggests estimate that the mantle is approximately $80\%$ MORB-source composition and the remaining $20\%$ is OIB-source composition \citep{Workman_and_Hart_2005, Arevalo_et_al_2009, Arevalo_et_al_2013}. Determining what proportion of OIB-source mantle is high-N and what proportion is MSL is difficult, but crucially important to the overall estimate of N in the mantle. As a first attempt, we assume that analyzed OIB represent a statistical sampling of the OIB-source mantle. There are 9 OIB samples with high N$_2$/$^{40}$Ar ($>10^3$) out of 61 total OIB samples. This corresponds to high-N being $\sim15\%$ of the OIB-source mantle, or $3\%$ ($15\%\times20\%$) of the total mantle. If we assume uniform distribution of $^{40}$Ar in the mantle, MSL ($97\%$ of the total) has $2.13\pm1.7\times10^{18}$ mol $^{40}$Ar and the high-N mantle ($3\%$ of the total) has $0.066\pm0.0054\times10^{18}$ mol $^{40}$Ar. Given N$_2$/$^{40}$Ar mentioned above ($120\pm11$ for MSL and $9.3\pm3.3\times10^3$ for high-N), we calculate N mass in the MSL and high-N mantle to be $7.2\pm5.9\times10^{18}$ and $17\pm15\times10^{18}$ kg N, respectively. Total mantle N is $24\pm16\times10^{18}$ kg, which is equivalent to $6\pm4$ ppm N. While there is uncertainty in this estimate, primarily related to the K concentration estimate and distribution in the mantle, we suggest that our calculation represents a lower estimate. A larger proportion  of high-N  mantle would significantly increase a N mass estimate.  

The most interesting and important aspect of our approach is the identification of two distinct mantle N reservoirs. The origin of both MSL and high-N components present a fascinating  geochemical quandry. MSL mantle has low N$_2/^{40}$Ar, low N concentration, but its  $\delta^{15}$N values describe two sub-populations: depleted $\delta^{15}$N  in MORB and enriched $\delta^{15}$N in OIB (Fig. \ref{fig:N2ArFull}). Given the low N concentration ($<1$ ppm) in most samples, the low N$_2/^{40}$Ar ratio should be caused by a high $^{40}$Ar content resulting from significant time since this material (and by proxy K) was at the surface of the Earth. Subducted material with variable N and K contents would require a long time to acquire enough $^{40}$Ar to push all samples towards a common trend. It therefore seems likely that MSL mantle is tapping a reservoir of older material derived from the surface. The $\delta^{15}$N values are interesting, as MORB values of $-5\permil$ are distinct from modern subducted material, which is around $+5~\textrm{to}~+7\permil$. OIB that fall along the MSL trend, however, show enriched $\delta^{15}$N values,  at $\sim5\permil$. The difference either means that MORB and OIB in the MSL are tapping N reservoirs of subducted material that are different in space \citep{Marty_and_Dauphas_2003} but not in time (i.e., both tap old material) or that the way N is processed in the MORB-and OIB-source mantle or eruptions is different. It is difficult to discern between these options at this time, though future modelling and experimental work would aid in this pursuit. 

The high-N mantle, in contrast, appears to be tapping relatively recently subducted surface material. This reservoir has high N$_2/^{40}$Ar, high N concentration, and enriched $\delta^{15}$N values in OIB and xenoliths (Fig. \ref{fig:N2ArFull}). High N concentration associated with high N$_2/^{40}$Ar implies this material is young, as it has not had sufficient time to accumulate $^{40}$Ar through radioactive decay \citep{Marty_and_Dauphas_2003}. The $\delta^{15}$N values are also very close to modern, oceanic sedimentary values, at $7.1\permil$. Overall, the high-N mantle appears to be young, N-rich, and received its N from subduction of surficial materials. 

In detail, there are differences between high-N OIB, OIB-xenoliths, and xenoliths. While both OIB basalts and OIB-xenoliths have relatively high N content (4.5 and 2.7 ppm), they have distinct $\delta^{15}$N values of $7.1\permil$ and  $-3.5\permil$, respectively.  Recall that the OIB-associated xenoliths are thought to represent the source rocks of coexisting OIBs. The difference in N isotopes could mean that N fractionates during partial melting, enriching the melt compared to the source. To our knowledge, there are no studies that quantify isotopic fractionation of N between partial melt and residual material in OIB genesis. If melt preferentially incorporates the heavy isotope, perhaps this could explain the observed relationship between OIB and their source. The remainder of the high-N xenolith population has enriched $\delta^{15}$N values of $4.5\pm2\permil$ and low N concentration of $0.35\pm0.07$ ppm. 

An alternate approach to explain the high N$_2/^{40}$Ar ratio of the high-N mantle would be some process whereby N is retained preferentially to K during subduction and recycling. Such a process would concentrate N more in the mantle than K, and therefore this material would have less $^{40}$Ar. It is possible that K is more mobile during subduction than N. There are synthesized NH$_4^+$-bearing micas (phengite), aluminosilicates (K-hollandite) \citep{Watenphul_et_al_2009} and clinopyroxenes \citep{Watenphul_et_al_2010} that are stable to eclogite-field conditions. As pyroxene is  more stable at greater depth in the mantle, it is possible that storage of N as NH$_4^+$ in pyroxene allows for it to be more effectively retained than K, whose host minerals (feldspars, micas) break down.  Further experimental work concerning N and K behaviour during subduction could help address this issue.   Discussed in some detail later, other locations that could fractionate N from K are the transition zone and lower mantle. In these reservoirs,  metallic Fe is stable and N may be retained in this metal, while K is not.  This is highly speculative, but further investigation of this high-N reservoir could help characterize the fate of volatiles in the mantle.

\subsubsection{Individual Mantle Domain Estimates}  \label{sec:indv}
An alternate approach to the Ar-based geochemistry is to attempt to break the mantle into different domains, and use measurements of xenoliths from those domains to estimate N mass. This approach may be more limited, due to relative paucity of analyzed samples as well as lack of material from the transition zone and lower mantle. Thus, we suggest the following be viewed as a minimum estimate. We will also only provide quantitative estimates for actual N content in domains that have been sampled, while for domains without direct samples we will discuss the capacity for N storage. 

We identify four sampled mantle reservoirs (Table \ref{tab:assumptions}): MORB-source, OIB-source, off-cratonic, and cratonic mantle. There are two reservoirs, the transition zone and lower mantle, that do not have N analyses from xenoliths (Fig. \ref{fig:mantlereservoirs}). Note this division is not intended to comment on chemical heterogeneity or stratification in the mantle, but merely to utilize different petrologic/geochemical proxies where appropriate to estimate the N content of the total mantle.

\begin{figure*}  
\centering
\includegraphics[keepaspectratio=true, width=\textwidth]{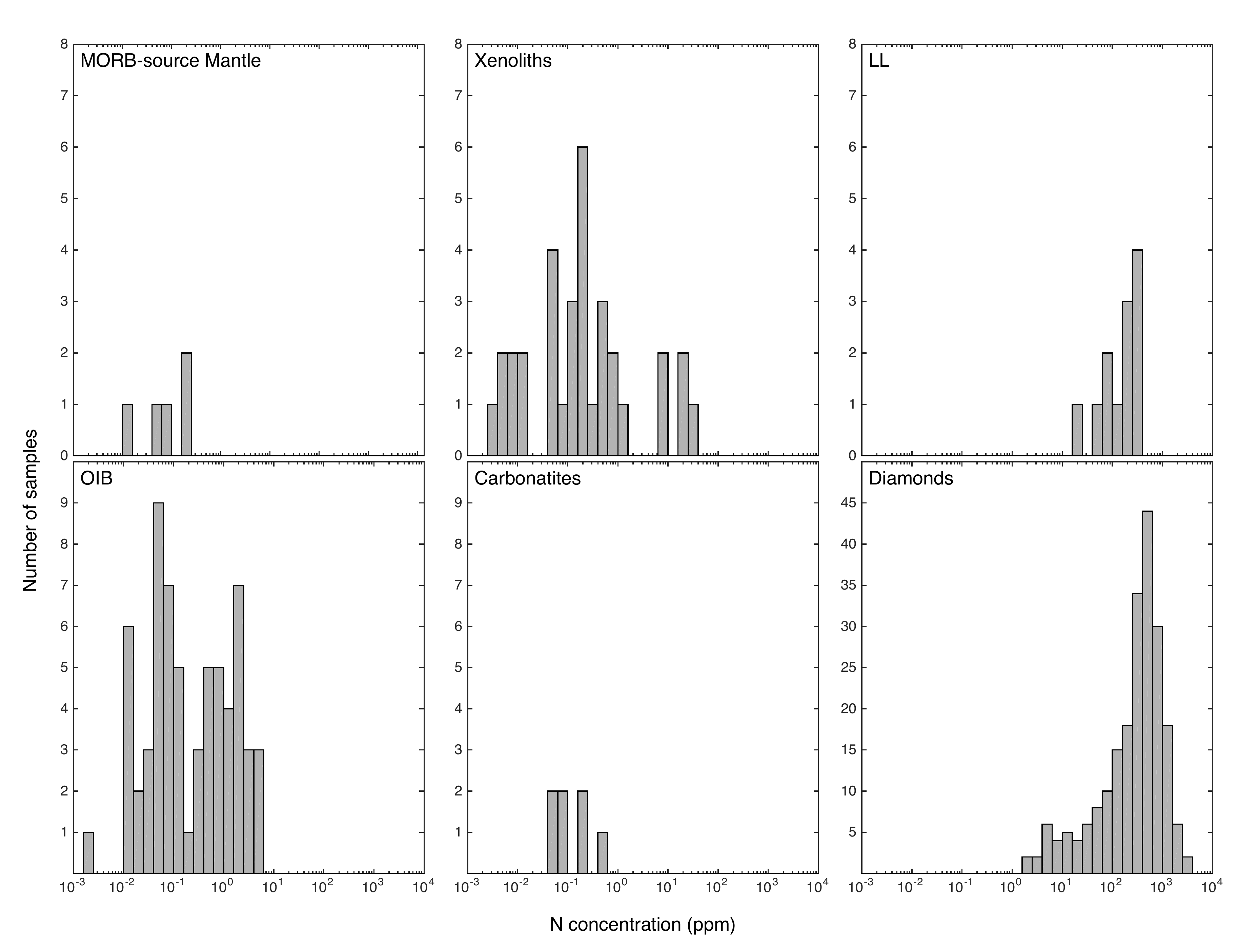}
\caption{Nitrogen concentrations in  mantle rocks, melts, and diamonds. While diamonds are well characterized, the number of analyses in other rock types are rare. MORB-source mantle samples include rocks from dredges and from ophiolites, while xenoliths are samples brought to the surface on continents. Lamproite and lamprophyre (LL) are thought to be sourced from hydrated mantle, and their high N concentration suggests effective recycling of N into this region of the mantle (Sec. \ref{sec:offcratonic}). Ocean Island Basalts (OIB) are discussed in detail in Sec. \ref{sec:NAr}, and carbonatites are presented simply as a comparison. See \cite{Dauphas_and_Marty_1999} for a discussion of these samples.  }
\label{fig:mantlehists}
\end{figure*}

\paragraph{MORB-source Mantle}
The amount of N in the MORB  mantle is largely a function of the efficiency of degassing during mantle partial melting and MORB genesis. Melting under oxic conditions seems to be efficient at removing N from source rocks into magma \citep{Libourel_et_al_2003}.  Our data compilation of MORB-source mantle rocks (peridotite, harzburgite) suggests  N content of $0.28\pm0.2$ ppm (Fig. \ref{fig:mantlehists}). Using the same MORB-source mass abundance from the previous section ($80\%$ of the mantle) yields a N content for the MORB-source  mantle of $0.74\pm0.1\times10^{18}$ kg N. Note that experimental \citep{Li_et_al_2013} and theoretical \citep{Mikhail_and_Sverjensky_2014} work suggest that in the middle to lowermost upper mantle redox, pressure, and pH conditions may be consistent with the presence of NH$_3$ or NH$_4^+$. These molecules may be retained more effectively than N$_2$, thus this portion of the mantle may be more N rich than indicated by xenoliths alone. 

\paragraph{OIB-source Mantle}
For this estimate, we turn to analyses of OIB as well as OIB-associated xenoliths as described in in section \ref{sec:NAr}. These rocks have an average N concentration of $0.7\pm0.5$ and $2.7\pm0.8$ ppm. Note that we include all OIB in this estimate, whereas in section \ref{sec:NAr} we identified two OIB-source reservoirs. In addition, OIB have likely experienced some degassing upon eruption, so this concentration should be viewed as a minimum. We assume that OIBs represent partial melts that melted under conditions conducive to quantitative N extraction from the source. Therefore, with  $10\%$ partial melt \citep{Winter_2001}, source concentration would be $0.07\pm0.04$ ppm N. This is much lower than the OIB-xenolith values, but again should be viewed as a minimum. Given an OIB-source proportion of $20\%$ yields a N mass of $0.06\pm0.04\times10^{18}$ kg for OIB based N concentration and $2.2\pm0.6\times10^{18}$ kg for OIB-xenolith based concentration. 

Additionally, while degassing during eruption has likely occurred, it is worth noting that the concentration of N in OIBs is of the same order of magnitude as the concentration of N in carbonatites from the Kola peninsula, which are around 0.11 ppm \citep{Dauphas_and_Marty_1999}. The carbonatites are thought to be sourced from fairly deep,  crystallized at depth, and to have experienced minimal degassing during emplacement. Carbonatite magmas, however, are likely sourced from a mantle domain distinct from the OIB-source mantle.

\paragraph{Off-cratonic Upper Mantle} \label{sec:offcratonic}
The sub-continental mantle can be broken into two domains: off-cratonic mantle, which has been influenced by Phanerozoic subduction and cratonic mantle, which is the stable mantle underneath cratons. We discuss the off-cratonic mantle first. 

Off-cratonic mantle  is roughly equivalent to the mantle wedge associated with subduction zones. Since mass balance studies suggest the majority of subducted N does not return to the atmosphere via arc magmatism \citep{Mitchell_et_al_2010, Busigny_et_al_2011}, it is possible that some of this N is retained in this reservoir. We invoke analyses of specific alkaline volcanic rocks,  lamprophyres and lamproites (LL), as proxy for mantle influenced by subduction. These rocks, though volumetrically small, are geographically widespread \citep{Winter_2001}, which indicates their potential as a useful proxy. 

Petrogenetic analysis of LL suggests that they  are sourced from hydrated mantle  composed of phlogopite (mica)-bearing harzburgite \citep{Tainton_and_McKenzie_1994,Mitchell_1995}, though some may be sourced from deeper in the mantle \citep{Murphy_et_al_2002}. 
Phlogopite harzburgite may be produced via a two step process: an initial mantle melting event, and the subsequent addition of fluids sourced from subducted continental/marine sediments. Later partial melting ($1-10\%$) of the  harzburgite produces LL magmas.   

A suite of LL from India  have N concentrations that range from 21-394 ppm \citep{Jia_et_al_2003}, with an average of $210\pm60$ ppm (Fig. \ref{fig:mantlehists}). The corresponding N content of the mantle source of LL depends, then, on the behaviour of NH$_4^+$ during melting. 

\begin{figure*}  
\centering
\includegraphics[keepaspectratio=true, width=0.8\textwidth]{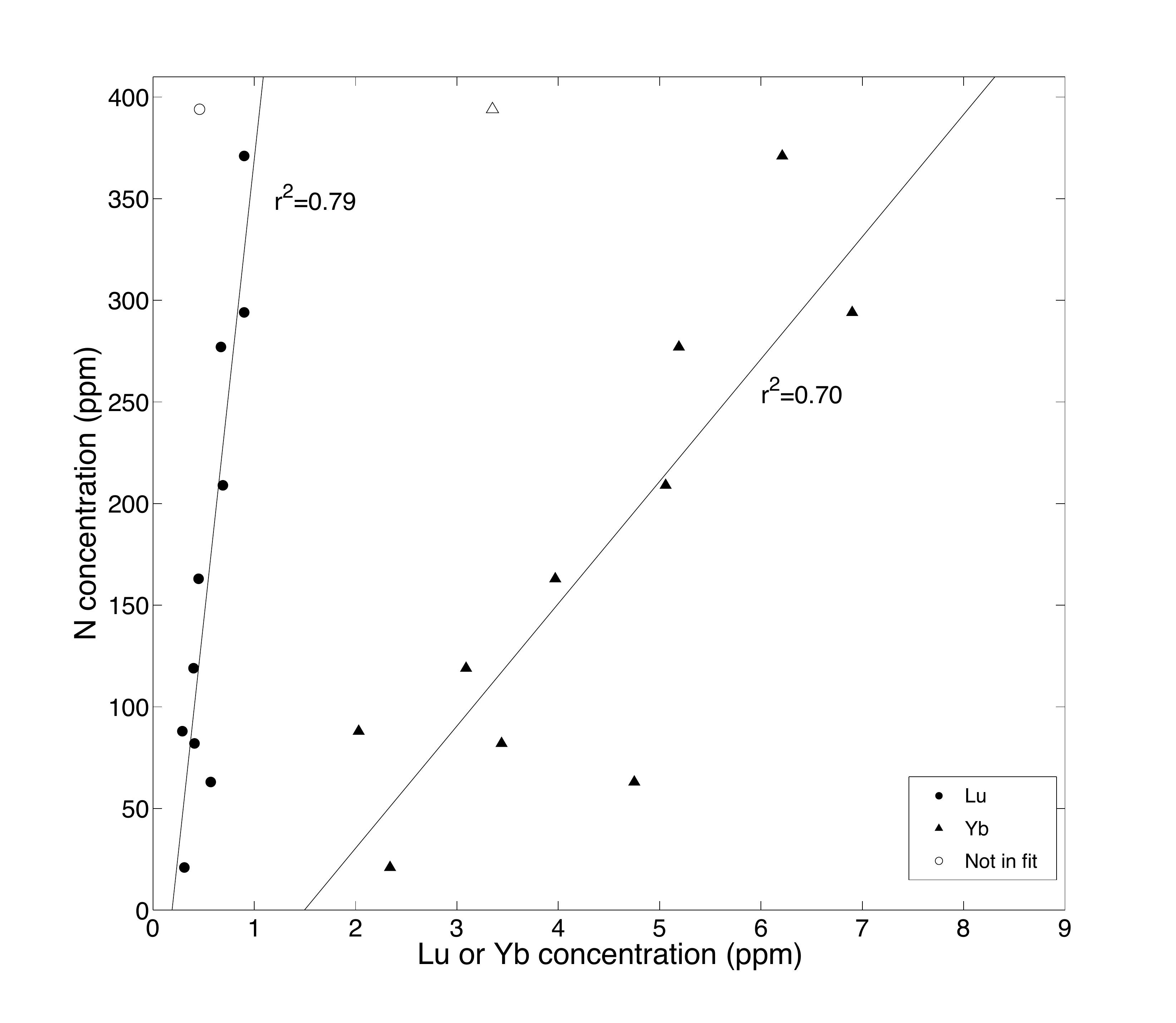}
\caption{Nitrogen and Lu or Yb concentration in lamproites/lamprophyres (LL) from \cite{Jia_et_al_2003}. Significant correlation suggests these elements behave similarly during melting, and we use this observation as an estimate for the N content of the off-cratonic upper mantle. Empty symbols were not included in regression.}
\label{fig:YbLu}
\end{figure*}

The Rare Earth Element (REE) profiles of the Indian LL may both constrain the compatibility of N in the source rock and could perhaps be used as a proxy for N in other samples where N was not measured explicitly. Ytterbium and Lu show a significant correlation with N (when disregarding a sample with high, 400 ppm, N), with r$^2$ values of 0.70 and 0.79, respectively (Fig. \ref{fig:YbLu}). This suggests that N, Yb, and Lu behave similarly during LL formation. 

The behaviour of Yb and Lu during LL formation is relatively well known, as K$_\textrm{D}$ values have been measured in minerals experimentally (Table \ref{tab:Kd}). We use these mineral K$_\textrm{D}$ values to calculate a bulk K$_\textrm{D}$ value, which is a simple weighted average, for a  phlogopite-harzburgite source rock with 60$\%$ olivine, $30\%$ pyroxene, and $10\%$ phlogopite. Bulk  K$_\textrm{D}$ are 0.0505--0.0979 for Yb  and 0.0636 for Lu \citep{Fujimaki_et_al_1984, Foley_and_Jenner_2004}. 

The K$_\textrm{D}$ of N has not been measured during LL formation, to our knowledge, so as a first approximation we will assume that it is equal to the K$_\textrm{D}$ of Yb or Lu, based on the strong correlation shown in Fig. \ref{fig:YbLu}. Using Equation \ref{eq:partial}, we calculate a N concentration in LL source of $35\pm7$ ppm based on Yb and $33\pm9$ ppm based on Lu for 10$\%$ partial melting. Assuming $1\%$ partial melting gives N concentration of $18\pm4$ ppm based on Yb and $15\pm5$ ppm based on Lu. Correspondingly, the mass of N would be  between $1.4\times10^{18}~\textrm{and}~3.4\times10^{18}$ kg.  

\begin{table*}

\caption{Partition coefficients of Yb and Lu in  lamproite/lamprophyre (LL). K$_\textrm{D}$ values shown are for a LL source rock that is $60\%$ olivine, $30\%$ pyroxene, and $10\%$ phlogopite. }
\centering
\begin{tabular}[h]{l  c c  c }  
\hline
&\multicolumn{2}{c}{\bf{K$_\textrm{D}$}}\\
Mineral	&	Yb	&	Lu &Reference			\\

Olivine	&	0.0091	&	0.0187	&\cite{Foley_and_Jenner_2004}\\	
Pyroxene	&	0.134--0.292	&	0.159&\cite{Foley_and_Jenner_2004}	\\
Phlogopite	&	0.0484	&	0.0471	&\cite{Fujimaki_et_al_1984}\\
Bulk rock &0.0505--0.0979 & 0.0636  \\

\hline

\label{tab:Kd}
\end{tabular}
\end{table*}

\begin{table*}

\caption{Nitrogen concentration and total mass estimates in the off-cratonic mantle based on analysis of lamproite/lamprophyre (LL) and K$_\textrm{D}$ values of Yb and Lu (Table \ref{tab:Kd}). Nitrogen behaviour is assumed to be similar to Yb and Lu (Fig. \ref{fig:YbLu}), and bulk K$_\textrm{D}$ values are then used to estimate N mass ($10^{18}$ kg)  using Eq. \ref{eq:partial}. We present estimates for 1 and 10$\%$ melt. }
\centering
\begin{tabular}[h]{l  c c  c c }  
\hline\\
&\multicolumn{2}{c}{N concentration (ppm)}&\multicolumn{2}{c}{N mass ($10^{18}$ kg)}\\
Melt&Yb-based&Lu-based&Yb-based&Lu-based\\

$10\%$&$35\pm7$&$33\pm9$&$3.4\pm0.7$&$3.2\pm0.8$\\\

$1\%$&$18\pm4$&$15\pm5$&$1.7\pm0.5$&$1.4\pm0.5$\\

\hline

\label{tab:LLnitrogen}
\end{tabular}
\end{table*}

\paragraph{Cratonic  Mantle}
We draw upon three data sources to calculate the N content of the cratonic upper mantle: xenoliths,  diamond analyses, and experimental petrology. Xenolith data gives rather different results than the estimate combining diamond and experimental data. We suggest that these approaches provide lower (xenolith) and upper (diamonds plus experiments) limits on the N content of the cratonic mantle. 

Xenolith N concentration is $0.28\pm0.2$ ppm. If these are representative of cratonic mantle, this reservoir has very low N mass, of $0.04\pm0.03\times10^{18}$ kg.  There are very few analyses of N in xenoliths, thus it is difficult to say how representative they are.

A second approach is to use the much more plentiful literature database for N analyses in diamonds. The majority of diamonds are formed in the cratonic lithosphere \citep{Cartigny_2005, Stachel_and_Harris_2009}, thus they should be able to reveal information about this reservoir. First, we calculate the average N content of diamonds to be $740\pm176$ (Fig. \ref{fig:mantlehists}). Diamonds form via precipitation from a fluid, so the next step is to estimate how much N remains in the fluid after diamond formation, and to assess the ability of the host rocks to absorb this extra N. A K$_\textrm{D}$ between N in diamond and N remaining in the diamond-forming fluid of 4 has been suggested based on study of the Jericho Kimberlite, in the Slave Craton of Canada \citep{Smart_et_al_2011}. Thus, given a K$_\textrm{D}$ of 4,  N concentration in residual diamond-forming fluid is $185\pm44$ ppm.

 Next, using experimental results, we calculate the capacity of the sub-cratonic lithosphere to retain N left in diamond-forming fluid after diamond formation (Eqs. \ref{eq:ol}--\ref{eq:pyx}). P-T-$f_{O_2}$ conditions appropriate to the sub-cratonic lithosphere are: T $=1400^{\circ}$ C, P$=6$ GPa, and $f_{O_2}=$FMQ-4 ($\Delta$NiNiO$=-2.3$) \citep{Frost_and_McCammon_2008}. Hence, we calculate a N capacity of 120 ppm for pyroxene and 5 ppm for olivine. If we assume a harzburgite-like composition (70$\%$ olivine, 30$\%$ pyroxene) \citep{Lee_et_al_2011}, a total N capacity of 40 ppm is calculated. If portions of the mantle are more fertile (i.e., higher pyroxene content) they would have correspondingly a higher N capacity. Analyses of N in natural olivine and pyroxene are much lower ($<1$ ppm, \cite{Yokochi_et_al_2009}) than this calculated upper limit, implying the actual content may be significantly lower than the capacity. 

The capacity of upper mantle minerals is consistent with relatively high ($\sim40$ ppm) N contents. We suggest that N remaining in diamond-forming fluid after diamond formation may be effectively sequestered in the sub-cratonic lithosphere. In addition, N concentration in the 10s of ppm is corroborated by estimates of C/N based on diamond analyses \citep{Cartigny_et_al_2001}. A N concentration of 40 ppm yields a N mass of $\sim5.6\times10^{18}$ kg. This is likely an upper estimate, as it assumes all cratonic mantle was infiltrated by diamond-forming fluids and that all N not included in diamonds was retained in the host rock. Uncertainty in the extent of diamond-forming fluid (if this is the source of N)  introduces error to this estimate, though if experimental relationships accurately describe the N solubility in upper mantle minerals there is no issue with storage capacity.

 Additional uncertainty may result from disagreement concerning the compatibility of N during diamond growth. An extensive compilation by \cite{Cartigny_et_al_2001} suggests that the incorporation of N into diamonds from diamond-forming fluids is a kinetic process: slow diamond growth results in low N content. Both measurements of Slave Craton diamonds \citep{Smart_et_al_2011} and synthetic diamonds  suggest that N behaves compatibly \citep{Stachel_and_Harris_2009}. The majority of diamonds with depleted $\delta^{13}$C values that are indicative of a recycled crustal input do not have commensurate enriched $\delta^{15}$N values \citep{Cartigny_2005}, which would be expected if the source of N and C was subducted oceanic material. This either indicates that recycled material that was incorporated into diamond growth had different $\delta^{15}$N values from modern surface reservoirs \citep{Marty_and_Dauphas_2003} or that C and N incorporation into diamonds is decoupled/depends only on growth rate \citep{Cartigny_et_al_2001}.

\paragraph{Transition Zone}

The transition zone (TZ) is the region in the mantle between 410 and 660 km depth, and is defined by mineral phase changes. At 410 km,  olivine changes its structure to the more tightly packed spinel-like crystal lattice of wadsleyite then ringwoodite. Below 660 km, another phase change occurs, and the spinel-structure mineral changes to bridgmanite, a high-pressure polymorph. These phase changes control redox chemistry, and the TZ is more reducing than the overlying upper mantle. The $f_{O_2}$ here is around IW-1, and both experiments and calculations suggest the presence of  0.1 wt.$\%$ metallic Fe \citep{Frost_and_McCammon_2008} in this region. 
 
Therefore, with 0.1-2 wt.$\%$ N dissolved \citep{Kadik_et_al_2011, Roskosz_et_al_2013} in 0.1 wt.$\%$ metallic Fe \citep{Frost_and_McCammon_2008}, N concentration would be 100--2000 ppm. This represents the N capacity of the transition zone, with a strong upper bound between $48\times10^{18}~\textrm{to}~960\times10^{18}$ kg N.  The high N potential of the transition zone, and indeed the lower mantle, was recently suggested based on the observation that the more deeply diamonds form, the less N they contain \citep{Smith_and_Kopylova_2014}. The authors suggest that the low-N diamonds grew in the presence of Fe-metal, which preferentially dissolved N. This is consistent with our literature review. 
 
Sequestering  N in the transition zone for significant periods of time may be difficult, however. Whole mantle circulation means that material in the transition zone does not stay there \citep[e.g.,][]{Nakagawa_and_Tackley_2012}. Both the upper mantle and lower mantle are more oxidizing that the TZ \citep{Frost_and_McCammon_2008}. Thus, when material moves out of the TZ, previously metal-bound N may be released into either fluids or minerals. This may preclude long-term N storage in the TZ. Without further evidence (petrologic or experimental), the transition zone remains a hypothetical reservoir for N. It is not considered in our total N estimates for the Earth.

\paragraph{Lower Mantle}

The lower mantle, which is defined by the phase transition at 660 km depth described above, is not represented by xenoliths or inclusions in our N database. We are unaware of any such analyses. As with the transition zone, we may only be able to speak to the storage capacity of this portion of the mantle. 

There is speculation that 1 wt.$\%$ metallic Fe may exist in the lower mantle \citep{Frost_et_al_2004}. High pressure/temperature experimental petrology demonstrates that N is quite soluble in metallic Fe at these conditions, with up to 8 wt.$\%$ \citep{Roskosz_et_al_2013}. Given a mantle with these proportions of Fe-metal and N solubility suggests a N capacity of $2.3\times10^{23}$ kg N. This value is orders of magnitude higher than any other reservoir in the planet, perhaps save the core. As with the transition zone, this is highly speculative, and would require further confirmation via geochemical or modelling study. 

If there are regions of the lower mantle that remain shielded from mantle convection and mixing, they may represent a location for storage of the Earth's ``missing N''. That is, the abundance of N compared to other volatiles in the BSE was previously estimated to be about an order of magnitude lower \citep{Marty_2012}. The lower mantle has more than enough capacity to store additional N. If, however, our Ar-based estimate for N abundance in the mantle is correct, there is no need to invoke hidden reservoirs of N, as we calculate  mantle N mass in line with other volatile abundances.

\begin{table*}
 \small
\caption{BSE N masses, shown in $10^{18}$ kg N. When adding  oceanic crust into totals, the range ($0.16\pm0.01~\textrm{to}~0.26\pm0.02$) was averaged to $0.21\pm0.01$. The approach column notates which method: literature compilation (LC) or Ar-based geochemistry (AR). The preferred value total is shown in bold. The AR approach for the mantle is preferred, because it more likely ``samples'' a greater extent of the mantle than xenoliths. Results are presented with comparison to \cite{Goldblatt_et_al_2009} (CG).}
\centering
\begin{tabular}[h]{ l  c c  c c}  
\hline\\

\bf{Reservoir}	&	\bf{Location}	&	\bf{This study}	& \bf{Approach} &	\bf{CG}		\\[2ex]
Oceanic Lithosphere	&	Sediments	&	0.41$\pm0.2$	 &LC&	$0.31\pm0.16$			\\
	&	Crust+Mantle	&	$0.16\pm0.01~\textrm{to}~0.26\pm0.02$	& LC&	$0.012\pm0.005$			\\[1ex]
Continental Crust	&	Igneous &		&	&$0.55\pm0.27$			\\
	&	Sedimentary	&		&&	$1.55\pm0.62$			\\
	& Total  & $1.7\pm0.1$ &LC& 2.1$\pm1.05$ \\[1ex]
Mantle	& MORB-Source  &$1.2\pm0.8$ & LC & \\
                 &OIB-source & $>0.06\pm0.04~\textrm{to}~2.2\pm0.6$ & LC & \\
                 & Off-cratonic & $1.4~\textrm{to}~3.4$ & LC&\\
                 &cratonic & $<5.6$ &LC&\\
	&transition zone, lower&$<100$ &&\\

	&	\emph{Total}	&	$>3.4\pm1.3-5.8\pm1.4$	&	LC&		\\[2ex]
Mantle	&	MORB-source-like Mantle	&	$7.2\pm5.9$	&AR &			\\
        &high-N &  $17\pm15$&AR&\\
	&\bf{\emph{Total}}& $\bf{24.2\pm16}$&AR&$\ge8.4\pm5.2$	\\[1ex]

  BSE Total               & LC Crusts and LC Mantle &$>5.7\pm1.3-8.1\pm1.4$ &&\\[1ex]
  \bf{BSE Total}	&	\bf{LC Crusts and AR Mantle}	&	\bf{27$\bf{\pm}$16}	&&	$10.8\pm5.3$		\\[2ex]
\hline

\label{tab:total}
\end{tabular}
\end{table*}

\section{Discussion}\label{sec:discussion}
 We find that our two methods for calculating the N budget of the Bulk Silicate Earth (BSE) are consistent. Both comparison to chondrite N abundance (``top-down'') and compilation of terrestrial analyses (``bottom-up'') of rocks and minerals suggest the BSE contains many atmospheric masses (PAN$=4\times10^{18}$ kg) of N.  The chondritic comparison suggests between $17\pm13\times10^{18}~\textrm{and}~31\pm24\times10^{18}$ kg N are in BSE while the terrestrial literature compilation suggests $27\pm16\times10^{18}$ kg N. Both estimates also have theoretical upper limits that are much higher (Table \ref{tab:total}), due to increased solubility of N in silicates at depth. This close agreement shows that our budget is internally self-consistent. It may also remove the concept of ``missing N'' \citep[e.g.,][]{Marty_2012, Halliday_2013}, as the mantle appears to have ample capacity for N sequestration.
 
 Importantly, our estimate is higher than previous estimates \citep[e.g.,][]{Goldblatt_et_al_2009}. The mantle appears to have a significant portion of the planetary N budget. High N content in the BSE has significant ramifications in relation to Earth and atmospheric geochemistry. 
  
\subsection{Key uncertainties}
Before discussing geochemical implications of the new budget presented herein, we touch briefly on the main uncertainties in our estimate. First, sparsely analyzed reservoirs (specifically the mantle and lower continental crust) inherently limit accuracy in estimates. Further analysis of these important reservoirs should be a focus of future work. Second, though the behaviour and partitioning of N during melting in the mantle is beginning to be tested experimentally \citep[e.g.,][]{Libourel_et_al_2003, Mysen_and_Fogel_2010, Li_et_al_2013, Li_et_al_2015}, the relative lack of studies necessarily introduces uncertainty. We suggest that N behaves similarly to Lu and Yb, though this relationship has not been assessed in all rock types. It is becoming clear that $f_{O_2}$, temperature, and pressure all exert strong control over N contents in the mantle. More experimental petrology and modelling studies would be valuable in elucidating the behaviour of N at the range of conditions found in the solid Earth. 

Thirdly, determining the geochemical character of the high-N mantle reservoir identified in Section \ref{sec:NAr} more accurately is of crucial importance. This reservoir, despite its small mass, may contain the majority of the N in the mantle. As it appears to be sampled by some OIB and xenoliths, more coupled N-Ar measurements of these rocks should help define this end member more completely. The stable isotopes are consistent with a recycled component, but the extent, residence time, and other geochemical properties are not fully constrained at this time.

\subsection{Evolution of the atmosphere-mantle system}
The atmosphere is not the main reservoir for N on Earth today. However, the processes responsible for the current distribution are not fully resolved. It remains ambiguous if the distribution between BSE and atmosphere has been the same as the current state, or if it has been different in the past. Nitrogen isotopic evidence and correlation with radiogenic $^{40}$Ar indicates that N derived from the surface of the Earth is subducted and cycled into the BSE \citep[e.g.,][]{Marty_1995, Marty_and_Dauphas_2003, Palya_et_al_2011}, so exchange has clearly occurred. The setting where this input occurs is subduction zones. At current subduction rates, $9.3\times10^{18}$ kg N could be subducted over 4 Ga of Earth history (Sec. \ref{sec:occrust}). So  at modern subduction rates, consistent since at least the Cretaceous \citep{Busigny_et_al_2011}, the entire atmosphere could be potentially pass through the mantle if  $\sim50\%$ of subducted N is recycled to the mantle (i.e., not returned to the atmosphere). This retention efficiency at modern subduction zones is poorly constrained \citep{Halama_et_al_2012}, with the Central American margin appearing to return significant sedimentary N ($100\%$) to the atmosphere \citep{Elkins_et_al_2006} while the colder Izu-Bonin-Mariana Arc retains most of subducted N ($\sim85\%$) to mantle depth \citep{Mitchell_et_al_2010}. As such,  there is no reason to rule out different efficiency in the past.

There are two reasons to suggest that N subduction was more efficient in the past. The first is that prior to the Great Oxidation Event, it is likely that the dominant  N ion in the oceans was NH$_4^+$ \citep{Garvin_et_al_2009}. Since NH$_4^+$ substitutes into sediments and oceanic lithosphere to enter the geologic cycle, a higher concentration might promote a greater N flux into subducting sediments and lithosphere. Indeed, an increase in N concentration is seen in Black Sea samples in the Quaternary, with sediments deposited under anoxic conditions having about twice as much N as sediments deposited under oxic conditions \citep{Quan_et_al_2013a}. Additionally, higher mantle temperature in the Archean \citep[e.g.,][]{Herzberg_et_al_2010} has been interpreted to lead to more vigorous mantle convection. Hotter mantle is also thought to produce thicker oceanic crust, due to greater degree of partial melting. Higher heat flow should cause extensive hydrothermal alteration, which could act as a sink for NH$_4^+$ from the ocean. A hotter mantle, however, may also be a drier mantle \citep{Korenaga_2011, Sandu_et_al_2011}, which would tend to slow convection and subduction. The interplay between these two factors, increased NH$_4^+$ in the ocean and  crust and hotter mantle, and any effects on N subduction are not known.

Progressive N sequestration over time  implies a more massive atmosphere in the Archean, which has potentially important effects on greenhouse warming \citep{Goldblatt_et_al_2009, Byrne_and_Goldblatt_2014}. Independent proxies for Archean paleopressure based on fossil raindrops \citep{Som_et_al_2012} and hydrothermal inclusions in quartz grains \citep{Marty_et_al_2013}, however, suggest the Archean atmosphere had the same amount (or less) of N as the modern. The raindrop constraint has subsequently been found to be too low; a more realistic constraint here is up to ten times modern density  \citep{Kavanagh_and_Goldblatt_2015}.  The N$_2$/$^{36}$Ar ratio from 3.0 Ga hydrothermal inclusions in quartz grains are approximately equal to the modern value \citep{Marty_et_al_2013}; this is inferred to suggest that the atmosphere may have had the same pressure (and therefore N content) as the modern day Earth. An earlier study on the same grains suggests, however, an upper limit for the N$_2$/$^{36}$Ar of 3.3 times the modern value \citep{Nishizawa_et_al_2007}. A robust empirical constraint on the amount of N$_2$ in the Archean atmosphere is enigmatic at this time.  Our work indicates substantial N is in the mantle, at least some of which has been recycled from the surface, so it is possible that the atmosphere was more massive in the past. Whether this indicates a monatonic drawdown or some more dynamic evolution of the atmosphere-mantle system is unknown at this time. 

The fate of subducted N has a direct effect on the $\delta^{15}$N value of the mantle. In fact, a significant missing piece of the N puzzle is the origin of the depleted $\delta^{15}$N signature of the MORB-source mantle, which exists in an apparent disequilibrium with isotopically enriched surface reservoirs.  As briefly discussed in Sec. \ref{sec:NAr}, there appear to be two classes of solutions to this dilemma: the MORB-source mantle records early subduction of depleted N \citep{Marty_and_Dauphas_2003} or that fractionations of N isotopes during deep Earth transport are responsible. A distinct MORB-only source mantle is not supported by our compilation herein, as many OIB have equivalent  N$_2/^{40}$Ar values as MORB. However,  MORB and OIB have different $\delta^{15}$N values at $-5\permil$ and $>0\permil$, respectively, so there must be some process to explain this distinction.  Possibly they represent pools of different subducted material that has ``aged'' the same amount to yield equivalent N$_2/^{40}$Ar. Preservation of distinct $\delta^{15}$N values implies incomplete mantle mixing over time.  The existence of the high-N mantle supports the existence of different mantle domains, though we cannot rule out that the different N$_2/^{40}$Ar signature of the high-N reservoir could be caused by  fractionation of N from during subduction or deeper mantle processing.  Redox reactions and possible N sequestration in the transition zone and lower mantle may all affect N geochemical signatures of mantle and mantle melts. 

It is becoming apparent through experimental \citep{Li_et_al_2013} and theoretical \citep{Mikhail_and_Sverjensky_2014} that NH$_4^+$ is the predominant species of N in much of the mantle. The geochemical behaviour of  NH$_4^+$ in subduction zones and  mantle reservoir materials (e.g., silicates, Fe-metal) should be a target for future research, as any isotopic fractionations are unknown to us at this time. 

\subsection{Bulk Earth $\delta^{15}$N and N delivery during accretion}
 A long-standing conundrum concerns the geochemistry and  isotopic signature of N delivered to the planet during accretion. The budget estimate based on CC and EC compositions assumes that significant N was present in the Earth during its early history to ensure its participation in core formation. The implication is that N was not delivered in any late veneer, but instead was delivered during the main phase of accretion. It must have been in a reduced form, either NH$_4^+$ or as nitride, and contained within either silicate lattices or Fe-metal, as N$_2$ would be too volatile, and perhaps not present in significant quantity in the inner solar system. Existing isotopic data is inconclusive on identifying a single meteoritic analogue. The presence of very depleted $\delta^{15}$N values from the mantle has been suggested to reflect preservation of primordial EC-like material, though these are analyses from diamonds \citep{Palot_et_al_2012}, and the behaviour and fractionation of N during diamond growth may not be fully understood. Additionally, CC have $\delta^{15}$N that is generally enriched. Our BSE+atmosphere bulk $\delta^{15}$N, given masses in Table \ref{tab:total} and $\delta^{15}$N values for MORB-source mantle ($-5\permil$), high-N mantle ($+5\permil$), continental crust ($7.3\permil$), and atmosphere ($0\permil$), is $2.1\permil$. 
 
 This estimate  is distinct from either CC ($\sim30\permil$) or EC($\sim-25\permil$), meaning either the Earth did not acquire its N from a single chondritic source or the process of core formation significantly fractionated N. The Bulk Earth $\delta^{15}$N value could be explained by a $\sim50\%$ contribution of CC-like and a $\sim50\%$ contribution of EC-like material during accretion, given no fractionation during core formation. If there was significant fractionation during core formation, it would have a large effect on residual N in the BSE. There is suggestion proposed that the dissolution of N into Fe-metal would preferentially fractionate light isotopes into the metal, following a Sievert's law-type reaction of N$_2\rightleftharpoons2$N dissolved \citep{Dauphas_and_Marty_1999}. This assumption suggests that breaking of the $^{14}$N--$^{14}$N bond is easier, so this isotope goes into the metal, leaving residual silicates enriched in $\delta^{15}$N. Were this the case, it would imply a higher contribution of EC to Earth's N. In principle, the same effect would be seen in NH$_4^+$, but to our knowledge, there are no experimental studies measuring N isotopes in coexisting metal and silicates. 
 
 A possible source of information concerning fractionation during core formation could be measurements of pallasites, which are meteorites thought to represent core-mantle boundaries of planetesimals. Measurements made by \cite{Prombo_and_Clayton_1993} on coexisting silicate and metal in pallasites show that the silicate phase is almost always isotopically enriched compared to the metal phase. The fractionation is up to $\sim70\permil$, which suggests fractionation during core formation could be quite large.  Experimental work at pressures appropriate to Earth's core formation could help  corroborate or quantify this effect for the Earth.

\section{Conclusions}
 We have compiled a nominal, self-consistent, whole-Earth N budget based on two independent approaches. Both a chondritic comparison and literature compilation of terrestrial analyses reveal the BSE contains many present atmospheric masses of N (PAN). Estimates are $17\pm13\times10^{18}$ kg to $31\pm24\times10^{18}$ kg N and  $27\pm16\times10^{18}$ kg N, respectively. Both estimates are  higher than previous work, and suggests we have found the supposed ``missing N''. Additionally, several conclusions are apparent from each approach. 
 
 The chondritic comparison is consistent with the Earth receiving its N during the main phase of accretion. This indicates significant  ($\sim10^{20}$ kg) N in the core, as N is siderophile under reducing conditions. If there is limited N-isotope fractionation during core formation, $\delta^{15}$N values for the BSE plus atmosphere suggest a mix of $\sim50\%$ enstatite-like and  $\sim50\%$ carbonaceous-like chondritic material can explain the N content of Earth. 

Our terrestrial literature compilation budget indicates that the continental crust ($\sim0.5$ PAN) and especially the mantle ($\sim6$ PAN) contain significant N. Interestingly, N-Ar and $\delta^{15}$N data from MORB, OIB, and xenoliths identifies the existence of two distinct N reservoirs: MORB-source like (MSL) and high-N. MSL, which is $\sim98\%$ of the mantle by mass, contains $\sim2$ atmospheric masses of N, has depleted $\delta^{15}$N, and its Ar-isotopes suggest material was subducted deep in the geologic past.  In contrast, high-N mantle has at least several atmospheric masses of N, enriched $\delta^{15}$N, and appears to have been subducted more recently. 

The presence of a large mass of subducted N in the mantle has important implications for the history of atmosphere-mantle communication over time. At present subduction rates, the entire atmospheric mass of N could be mixed into the mantle if only $\sim50\%$ of down-going N is returned to the atmosphere via arc volcanism. Nitrogen-Ar systematics indicate that the atmosphere and MSL are well mixed, and therefore that the mantle may serve to buffer the amount of N in the atmosphere. More reduced geochemical conditions at the surface and hotter mantle temperatures in the Archean may have lead to more efficient N subduction in the past, perhaps indicating a more massive atmosphere early in Earth history that has been progressively sequestered into the mantle. 

This is an exciting time for research concerning the geologic N cycle. While the overall cycle is understood, there are areas for future research that are critical for more fully understanding N in the solid Earth. More analyses of N and Ar in OIB and xenoliths would help clarify the nature and extent of the high-N mantle. Experimental work investigating the behaviour of N, specifically as NH$_4^+$, during subduction and under mantle conditions should help reveal geochemical and isotopic fractionations during mantle transport. Modelling work, anchored to the budget presented herein, can elucidate the interchange of N and other surface materials through the solid Earth over geologic time.

\section*{Acknowledgements}
The authors would like to acknowledge Dante Canil, Rameses D'Sousza, and Brendan Byrne for constructive feedback and discussion concerning this manuscript. We also thank  Ralf Halama and Yuan Li for careful review of the manuscript. Funding was provided by NSERC Discovery grant to CG.

\end{document}